\documentclass{article}

\usepackage{amsmath,amsthm,amssymb}
\usepackage{mathtools}
\usepackage{hyperref}
\usepackage{graphicx}
\usepackage{algorithm}
\usepackage{algpseudocode}
\usepackage{booktabs}
\usepackage{caption}
\usepackage{comment}
\usepackage{fullpage}
\usepackage[T1]{fontenc}
\usepackage[utf8]{inputenc}
\usepackage{threeparttable}
\usepackage{setspace}
\usepackage{placeins}
\usepackage{natbib}
\usepackage{sansmath}
\usepackage{csquotes}
\usepackage{scalerel}
\usepackage{bm}
\usepackage{fancyhdr}
\usepackage{geometry}

\DeclareMathAlphabet{\mathbfsf}{OT1}{cmss}{bx}{n}

\newcommand{\Beta}{\mathrm{B}}

\begin{document}

\pagestyle{fancy}

\fancyhead{} 
\fancyhead[L]{In production at Multivariate Behavioral Research} 
\fancyhead[R]{Kavelaars et al.} 

\title{Bayesian multivariate logistic regression for superiority and inferiority decision-making under observable treatment heterogeneity}

\author{
	Xynthia Kavelaars\textsuperscript{a,b}\thanks{Contact: xynthia.kavelaars@ou.nl\\
A version of this paper is in production at Multivariate Behavioral Research}, 
	Joris Mulder\textsuperscript{a}, and 
	Maurits Kaptein \textsuperscript{c}}

\date{\footnotesize{\textsuperscript{a} Department of Methodology and Statistics, Tilburg University, The Netherlands}\\
	\footnotesize{\textsuperscript{b} Department of Theory, Methodology and Statistics, Open Universiteit, The Netherlands\\
		\footnotesize{\textsuperscript{c}Eindhoven University of Technology, Mathematics and Computer Science, The Netherlands}}
}

\maketitle

\begin{abstract}
	
	The effects of treatments may differ between persons with different characteristics.
	Addressing such treatment heterogeneity is crucial to investigate whether a patient with specific characteristics is likely to benefit from a new treatment.
	The current paper presents a novel Bayesian method for superiority and inferiority decision-making in the context of randomized controlled trials with multivariate binary responses and heterogeneous treatment effects.
	The framework is based on three elements: a) Bayesian multivariate logistic regression analysis with a P\'olya-Gamma expansion for model fitting; b) a transformation procedure to 
	transfer obtained regression coefficients to a more intuitive multivariate probability scale (i.e., success probabilities and the differences between them); and c) a compatible decision procedure for treatment comparison with prespecified decision error rates. 
	Procedures for a priori sample size estimation under a non-informative prior distribution are included. 
	A numerical evaluation demonstrated that decisions based on a priori sample size estimation resulted in anticipated error rates among the trial population as well as subpopulations and individual patients. 
	Further, average and conditional average treatment effect parameters could be estimated unbiasedly when the sample was large enough.
	Illustration with the International Stroke Trial dataset demonstrated that stroke patients with different blood pressure values react differently to the treatment: Something that would have remained undetected when analyses were limited to average treatment effects.

\end{abstract}


\section{Introduction}\label{s:intro}

The current paper focuses on estimating treatment effects among populations, subpopulations, and individual patients in the context of two-arm randomized controlled trials (RCTs) with multiple (correlated) binary dependent variables.
Such RCTs are randomized experiments with subjects being assigned at random to either an experimental or a control group, often having the objectives a) to evaluate whether an experimental treatment is superior or inferior to a control condition; and b) to inform the prescription of treatments to patients in (clinical) practice \citep{FDA2016}.
Although RCTs are broadly applicable to experimental research in general, we focus on the health domain and use the word \enquote{treatment} to refer to psychological and medical interventions in the broad sense.
These interventions include - but are not limited to - behavioral therapies, pharmacological support, and other experimental types of care.

Such RCTs often assess multiple types of (clinical) events (e.g., quitting substance abuse, death), functional measures (e.g., memory decline, ability to walk), or disease symptoms (e.g., fatigue, anxiety) \citep{FDA2017}.
Studying multiple dependent variables in RCTs is useful, since multiple dependent variables provide multidimensional insights into the effects of a treatment and since analyzing multiple dependent variables together has the potential to improve the connection between clinical and statistical desicion-making.
More specifically, multiple effects of the intervention can be combined and weighted in various ways to provide a single statistical decision regarding superiority or inferiority, similar to decisions regarding treatment prescription made by therapists or clinicians \cite[e.g.,][]{Pocock1987,OBrien1984,Murray2016}.
Whereas performing multiple univariate analyses on individual dependent variables is a common strategy to deal with data from multiple dependent variables, a single multivariate analysis is often preferable from a statistical point of view \citep{Senn2007,Ristl2018,FDA2017,Murray2016}.
Multivariate analysis takes the correlation between dependent variables into account and therefore has the potential to reduce decision errors: 
Correlations influence the sample sizes required for decision-making with prespecified error rates and provoke under- or overpowerment when falsely omitted \citep{Chow2017,Sozu2010,Xiong2005}.

RCTs often focus on average treatment effects (ATEs) among the study population when comparing interventions \citep{Thall2020}.
Average treatment effects can be sufficiently insightful when the effects of a treatment are relatively homogeneous over the trial population. 
In this case, patients react relatively similarly to the treatment.
However, average effects may give a limited, or even erroneous, impression when the actual effects of a treatment are heterogeneous and thus interact with characteristics of patients.  
In that case, patients differ in their reactions to the treatment. 
Taking characteristics of patients into account in the estimation of treatment effects (i.e., estimating conditional average treatment effects; CATEs) can then contribute to a better understanding of the treatment's potential for an individual patient.
Despite efforts to provide statistical methodology to model CATEs \citep[e.g.,][]{Wang2015,Yang2021,Jones2011}, investigating these effects is not the standard yet: Thall noted that "the great majority of clinical trial designs ignore the possibility of treatment-covariate interactions, and often ignore patient heterogeneity entirely" \citep[p.1]{Thall2020}. 
This is unfortunate as addressing conditional effects in the evaluation of treatments is crucial to a) identify how likely a specific patient will benefit from a treatment; and b) optimize treatment results of individual patients via personalized treatment assignment \citep{Goldberger2013,Hamburg2010,Wang2015,Simon2010}.

An example of a trial with multiple dependent variables and potential treatment heterogeneity is the International Stroke Trial \citep[IST;][]{Sandercock2011,ISTCG1997}.
Strokes may have far-reaching implications for the quality of life, as they may be recurring and/or lead to long-term impaired (daily) functioning. 
The IST investigated whether the short-term and long-term perspective of stroke patients can be improved with anti-thrombotic drug therapy.
The average treatment differences in the IST were small, so one might conclude that treatment with one of these drugs was marginally effective.
However, these overall findings were based on the assumption that specific characteristics of patients (e.g., sex or age) and/or disease (e.g., type of stroke or functional status after stroke) did not interact with the treatment to produce different effects for different patients.
Average treatment effects could, for example, not reveal whether older patients have better prospects in terms of short-term damage risk and/or long-term recovery potential than younger patients.  
Clearly, hypothetical heterogeneous effects as these would have clinically and psychologically relevant implications and advocate the development of more personalized treatment policies.

{
	While multivariate treatment effects for patients with specific characteristics are theoretically relevant for many contemporary RCTs contributing to the personalization of treatments, decision-making under treatment heterogeneity in the multivariate context is considerably more complex compared to the non-heterogeneous and/or univariate setting. %
	Generalizations to the heterogeneous and multivariate context are subject to assumptions that need to be carefully evaluated in light of the research problem at hand. %
	First, the multivariate setting demands an analysis method that incorporates the correlation between dependent variables (i.e., a multivariate analysis method) to obtain accurate decision error rates \citep[e.g.,][]{Sozu2010,Sozu2016,Kavelaars2020}.
	Ignoring or misspecifying a non-zero correlation can result in over- or underestimation of the required sample size and thus affects the statistical power of the analysis. %
	For accurate inference regarding conditional average treatment effects, the analysis should not only include the overall correlation among the trial population, but should also be flexible enough to deal with correlations that differ over subpopulations. %
	The latter is not evident in existing multivariate analysis methods for binary dependent variables: Some methods impose the marginal correlation structure of the trial population on subpopulations (e.g., multivariate probit models by  
	\citet{Chib1995} or 
	\citet{Rossi2005} and multivariate logit models by 
	\citet{Malik1973} and 
	\citet{OBrien2004}). %
	Second, the interpretation of treatment effects can be complex in multivariate non-linear models. %
	Creating insights into so-called marginal effects (i.e., treatment effects on the individual dependent variables) is recommended in treatment comparison, demanding any multivariate method to return interpretable univariate effects \citep{FDA2017, OBrien2004}. %
	Some existing multivariate models lack insight into marginal distributions \citep[e.g][]{Malik1973}. %
	Third, some multivariate methods estimate a single regression parameter to capture the relation between a covariate and all dependent variables \citep[e.g.,][]{OBrien2004,Rossi2005}. %
	The latter assumes that all dependent variables vary identically over the full support of the covariate. %
	In other words, all relations between the covariate and the outcome variable have the same size and direction. %
	Clearly, such an assumption may be too strict to hold in practice. %
	\label{response:alternatives}}

{
	In order to deal with the complexity of heterogeneous, multivariate treatment effects, we build upon an existing Bayesian multivariate Bernoulli framework for superiority decision-making proposed by 
	\citet{Kavelaars2020}. %
	The existing procedure consists of three major components: a) a multivariate analysis model to estimate unknown parameters; b) a transformation procedure to interpret treatment effects on the (more intuitive) probability scale; and c) a compatible decision procedure to make inferences regarding treatment superiority with prespecified error rates. %
	The analysis procedure has advantages over several other approaches, as it relies on a multinomial distribution and therefore has the flexibility to model univariate effects and correlations between dependent variables. %
	The transformation procedure facilitates the interpretation of treatment comparison: marginal (i.e., univariate) probabilities, multivariate probabilities, and differences between (multivariate) probabilities are can be used in inference as well. %
	The decision procedure is suitable for Bayesian inference and can be flexibly applied with several decision rules for multiple dependent variables. %
	Noteworthy is a decision rule with a compensatory mechanism, that can weigh dependent variables by their importance and has a natural compensatory mechanism that can balance positive and negative treatment effects. %
	{
		With this decision procedure, decisions regarding treatment superiority can be made with targeted decision error rates (i.e., Type I and Type II errors) and a priori computed sample sizes. %
	}\label{response:sample_size}

	\citet{Kavelaars2020} proposed a multivariate Bernoulli model for multivariate Bernoulli outcomes to estimate average treatment effects and to make decisions based on multivariate treatmeat effects. %
	In the current paper we propose a more flexible modeling framework for multivariate Bernoulli outcomes using Bayesian multivariate logistic regression models. %
	This extension allows us to model and estimate multivariate treatment effects for different (sub)populations based on available covariate information. %
	Moreover, to make decisions about multivariate treatment effects for these different subpopulations, we extend the decision procedure of Kavelaars et al. \citet{Kavelaars2020} to the new multivariate logistic regression model. %
	Additionally, sample size recommendations are provided for estimating and decision-making under this framework.
	\label{response:framework}}

{
	Note that the proposed multivariate modeling framework aims to estimate heterogeneous multivariate treatment effects that are caused by observed covariate information and to make decisions about treatment superiority. %
	Thereby, the aim is different from mixture modeling which aims to capture unobserved (treatment) heterogeneity using latent variables (either discrete or continuous). %
	Mixture models use response data to cluster respondents based on their patterns of outcome data (e.g., patterns of symptoms), where each cluster has an individual distribution that forms a constituent of the mixture \citep{McLachlan2019}. %
	The proposed regression model does not include latent variables (either discrete or continuous) to capture unobserved heterogeneity. %
	Instead, multivariate (logistic) regression uses observed covariate information to define patient groups of interest, often based on theoretical (such as accepted cutoff values for high and low blood pressure) or statistical (such as those respondents with more extreme scores than one standard deviation below or above the mean) grounds. %
	Subgroups are thus bounded by criteria specified by the researcher, rather than by response patterns in the data. %
	\label{response:mixture}}

The paper is organized as follows. 
In the next section, we introduce the decision framework, including the multivariate logistic regression model to obtain a sample from the multivariate posterior distribution of regression coefficients, a transformation procedure to find posterior treatment differences, and a decision procedure to draw conclusions regarding treatment superiority and inferiority. 
The section on capturing heterogeneity explains how the framework can be applied to different patient populations.
We evaluate frequentist operating characteristics of the framework via simulation in the numerical evaluation section. 
Next, we illustrate the methods with data from the International Stroke Trial and conclude the paper with a discussion.

\section{Decision-framework}\label{s:model}

\subsection{Multivariate logistic regression}\label{ss:logistic_regression}

Response $y^{k}_{i}$ is the binary response for subject $i$ on outcome variable $k \in \{1,\dots,K\}$, where $y^{k}_{i} \in \{0,1\}$, $0=$ failure and $1=$ success. 
Vector $\bm{y}_{i} = (y^{1}_{i},\dots,y^{K}_{i})$ is the multivariate (or joint) binary response vector of subject $i$ on $K$ dependent variables and has configuration $\bm{H}_{q\cdot}$, which is one of the $Q = 2^{K}$ possible response combinations of length $K$ given in the $q^{th}$ row of matrix $\bm{H}$: 
\begin{flalign}
	\bm{H} = & 
	\begin{bsmallmatrix}
		1 & 1 & \dots & 1 & 1 \\
		1 &1 & \dots & 1 & 0 \\
		& & \dots & &\\
		0 & 0 & \dots & 0 & 1 \\
		0 & 0 & \dots & 0 & 0\\
	\end{bsmallmatrix}
\end{flalign}
The probability of $\bm{y}_{i}$ can be expressed in two meaningful and related ways.
First, $\bm{\theta}_{i} = (\theta^{1}_{i},\dots,\theta^{K}_{i})$ denotes the vector of $K$-variate success probabilities on individual outcome $1,\dots,K$, where $\theta^{k}_{i} = p(y^{k}_{i} = 1)$. 
Second, $\bm{\phi}_{i} = (\phi^{1}_{i},\dots,\phi^{Q}_{i})$ denotes the vector of $Q$-variate joint response probabilities, where $\phi^{q}_{i} = p(\bm{y}_{i} = \bm{H}_{q\cdot})$ 
and sums to unity.
The joint response of subject $i$ can be conditioned on covariates in vector $\bm{x}_{i} = (x_{i1},\dots,x_{iP})$.
In this case, the probabilities of response vector $\bm{y}_{i}|\bm{x}_{i}$ are expressed as functions of $\bm{x}_{i}$, namely $\bm{\phi}_{i}(\bm{x}_{i})$ and $\bm{\theta}_{i}(\bm{x}_{i})$.

Joint response probability $\phi^{q}_{i}(\bm{x}_{i})$ maps the dependency of joint response probabilities on covariates $\bm{x}_{i}$ via a multinomial logistic function:
\begin{flalign}\label{eq:psi2phi}
	\phi^{q}_{i}(\bm{x}_{i}) 
	= & \frac{
		\text{exp} \left[\psi^{q}_{i}(\bm{x}_{i})\right]}
	{\displaystyle\sum_{r=1}^{Q-1} \text{exp} \left[\psi^{r}_{i} (\bm{x}_{i}) \right] + 1}
\end{flalign}
\noindent for response categories $q = 1,\dots,Q-1$.
In Equation \ref{eq:psi2phi}, $\psi^{q}_{i}(\bm{x}_{i})$ reflects the linear predictor of response category $q$ and subject $i$:
\begin{flalign}\label{eq:psi}
	\psi^{q}_{i} (\bm{x}_{i}) = &\beta^{q}_{0} + \beta^{q}_{1} x_{i1} + \dots + \beta^{q}_{P} x_{iP}.
\end{flalign} 
\noindent Here, $x_{ip}$ can be a treatment indicator, a patient characteristic, or an interaction between these.
Vector $\bm{\beta}^{q} = (\beta^{q}_{0}, \beta^{q}_{1}, \dots, \beta^{q}_{P})$ is the vector of regression coefficients of response category $q$.
To ensure identifiability, all regression coefficients of response category $Q$ are fixed at zero, i.e., $\bm{\beta}^{Q} = \bm{0}$.

The likelihood of response data follows from taking the product over $n$ individual joint response probabilities from Equation \ref{eq:psi2phi} of $Q$ response categories:
\begin{flalign}\label{eq:NH_mult_lik_beta}
	l(\bm{y}|\bm{\beta}, \bm{x}) 
	= & 
	\prod_{i=1}^{n} 
	\prod_{q=1}^{Q-1}
	\left(
	\frac{\text{exp} \left[ \psi^{q}_{i} (\bm{x}_{i}) \right]}
	{\displaystyle\sum_{r=1}^{Q-1} \text{exp} \left[ \psi^{r}_{i} (\bm{x}_{i}) \right] + 1 }\right)^{I(\bm{y}_{i}=\bm{H}_{q\cdot})}
	\left(
	\frac{1}
	{\displaystyle\sum_{r=1}^{Q-1} \text{exp} \left[ \psi^{r}_{i} (\bm{x}_{i}) \right] + 1 }\right)^{I(\bm{y}_{i}=\bm{H}_{Q\cdot})}
	.
\end{flalign}
Bayesian analysis is done via the posterior distribution which is given by
\begin{flalign}\label{eq:posterior}
	p(\bm{\beta}^{q}, \bm{\beta}^{q}, \bm{\Sigma}^{q} | \bm{y}) \propto &
	p(\bm{y}|\bm{\beta}^{q}) p(\bm{\beta}^{q} | \bm{\beta}^{q}, \bm{\Sigma}^{q}) p(\bm{\beta}^{q}) p(\bm{\Sigma}^{q}), 
\end{flalign}%

\noindent where $p(\bm{\beta}^{q})$ reflects the prior distribution of the unknown parameters before observing the data. 
Posterior sampling can be done with a Gibbs sampling algorithm based on a P\'olya-Gamma expansion \citep{polson2013}. 
Computational details of this procedure can be found in Appendix \ref{app:posterior_computation}.

\subsection{Transformation to treatment differences}\label{ss:transformation}

In contrast to several other regression analyses, the obtained multinomial regression coefficients have no straightforward interpretation.
We aim to make the posterior sample of regression coefficients interpretable in terms of a treatment difference, which is defined as the (multivariate) difference between success probabilities of two treatments. 
To this end, we execute the following multistep procedure with a fictive setup of the IST trial as running example.

Suppose we are interested in the effect of a combined drug therapy (Heparin plus Asparin; $T_{H+A}$) vs. single drug therapy (Aspirin only; $T_{A}$) on recurrent stroke on the short-term ($y^{strk}$) and dependency on the long-term ($y^{dep}$). 
There is a total of $Q=4$ response categories: $\{y^{strk}=1, y^{dep}=1\}$, $\{y^{strk}=1, y^{dep}=0\}$, $\{y^{strk}=0, y^{dep}=1\}$, $\{y^{strk}=0, y^{dep}=0\}$, which we refer to as $\{11\}$, $\{10\}$,$\{01\}$, and $\{00\}$ respectively.
The treatments are blood thinning agents and may thus interact with the patient's blood pressure.
Therefore, we include systolic blood pressure at the time of randomization as a covariate, so that we can estimate conditional effects for patients with different values of blood pressure, resulting in the following model: 
\begin{flalign}\label{eq:psi_IST2}
	\psi^{q}_{i}(\bm{x}_{i}) = & \beta^{q}_{0} + \beta^{q}_{1} T_{i} + \beta^{q}_{2} bp_{i} + \beta^{q}_{2} bp_{i} T_{i},
\end{flalign}
where $\bm{x}_{i} = (T_{i}, bp_{i}, bp_{i}T_{i})$. 
The transformation procedure is then as follows:
\begin{enumerate}
	\item 
	\textbf{Regression coefficients $\bm{\beta}$ to joint response probabilities $\bm{\phi}_{T}(\bm{x})$:}

	In the first step, the posterior sample of regression coefficients $\bm{\beta}$ is transformed to a treatment effect in terms of joint response probabilities $\phi_{Ti}(\bm{x}_{i})$ for each treatment $T \in \{0,1\}$.
	Linear predictor $\psi^{q}_{i}(\bm{x}_{i})$ is then transformed to individual joint response probability $\phi^{q}_{i}(\bm{x}_{i})$ via the multinomial logistic function in Equation \ref{eq:psi2phi}:
	\begin{flalign}\tag{\ref{eq:psi2phi} revisited}
		\phi^{q}_{i}(\bm{x}_{i}) 
		= & \frac{
			\text{exp} \left[\psi^{q}_{i}(\bm{x}_{i})\right]}
		{\displaystyle\sum_{r=1}^{Q-1} \text{exp} \left[\psi^{r}_{i} (\bm{x}_{i}) \right] + 1}
	\end{flalign} 
	For example, the probability that patient $i$ in the IST does not experience a new stroke and is dependent after six months can be expressed as:
	\begin{flalign}\label{eq:phi_app}
		\phi^{3}_{T_{i}} (\bm{x}_{i}) & = p(\bm{y}_{i} (\bm{x}_{i}) = \{01\}) \\\nonumber
		& = \frac{
			\text{exp} \left[\psi^{3}_{i}(\bm{x}_{i})\right]}
		{\displaystyle\sum_{r=1}^{Q-1} \text{exp} \left[\psi^{r}_{i} (\bm{x}_{i}) \right] + 1}.\nonumber
	\end{flalign}
	This probability can be computed for the other joint response combinations as well. 
	Note that we are in fact interested in joint response probability $\bm{\phi}_{T} (\bm{x})$, which reflects a treatment effect among a (sub)population defined by $\bm{x}$.
	This notation is more general than the joint response probability of an individual patient with covariates $\bm{x}_{i}$.  
	The population can be reflected by an individual patient (e.g., with a systolic blood pressure of $100$) in some situations, while other cases target the entire study population (e.g., no restriction on systolic blood pressure) or a subpopulation of interest (e.g., with a systolic blood pressure above $150$). 
	These variations have slightly different computational procedures, which we discuss in more detail in Section \ref{s:heterogeneity}.

	\item 
	\textbf{Joint response probabilities $\bm{\phi}_{T}(\bm{x})$ to multivariate success probabilities $\bm{\theta}_{T}(\bm{x})$:}
	
	The next step in the transformation involves the conversion from joint response probabilities $\bm{\phi}_{T} (\bm{x})$ to multivariate success probabilities of individual dependent variables $\bm{\theta}_{T} (\bm{x})$. 
	Especially when the number of dependent variables increases, success probabilities are more straightforward in their interpretation than joint response probabilities.

	The relation between both quantities is additive: 
	Success probability $\theta^{k}_{T}$ on outcome $k$ and treatment $T$ equals the sum of a selection of elements of $\bm{\phi}_{T}$, denoted by matrix $\bm{U}_{k}$: 
	\begin{flalign}\label{eq:phi2theta}
		\theta^{k}_{T} (\bm{x})
		& = \displaystyle\sum_{q=1}^{Q} \phi^{q}_{T} (\bm{x}) I(\bm{H}_{q\cdot} \in \bm{U}_{k}). 
	\end{flalign}
	\noindent
	Selection $\bm{U}_{k}$ consists of the $2^{K-1}$ rows of $\bm{H}$ that have their $k^{th}$ element equal to $1$.
	More concretely, the two dependent variables from the IST  
	are the following combinations,  where we drop the dependency on $\bm{x}$ for notational simplicity.
	$$\bm{H} =  
	\begin{bsmallmatrix}
		1 & 1 \\
		1 & 0 \\
		0 & 1 \\
		0 & 0 \\
	\end{bsmallmatrix}\text{, } 
	\bm{U}_{strk} =  
	\begin{bsmallmatrix}
		1 & 1 \\
		1 & 0 \\
	\end{bsmallmatrix} \text{, and }
	\bm{U}_{dep} =  
	\begin{bsmallmatrix}
		1 & 1 \\
		0 & 1 \\
	\end{bsmallmatrix}.$$
	Hence, the multivariate success probabilities in $\bm{\theta}_{T} = (\theta^{strk}_{T}, \theta^{dep}_{T})$ consists of univariate success probabilities:
	\begin{flalign}
		\theta^{strk}_{T} & = 
		p(\bm{y}_{i} (\bm{x}_{i}) = \{11\}) + \
		p(\bm{y}_{i} (\bm{x}_{i}) = \{10\}) \\\nonumber 
		& = \phi^{1}_{T} + \phi^{2}_{T} &&\\\nonumber
		\theta^{dep}_{T} & = 
		p(\bm{y}_{i} (\bm{x}_{i}) = \{11\}) + \
		p(\bm{y}_{i} (\bm{x}_{i}) = \{01\}) \\\nonumber
		& =  \phi^{1}_{T} + \phi^{3}_{T}.&& \nonumber
	\end{flalign} 
	The correlation between these dependent variables is captured in joint response probabilities $\bm{\phi}_{T} (\bm{x})$ and automatically taken into account in further transformations \citep{Olkin2015,Dai2013}.
	
	\item
	\textbf{Success probabilities $\bm{\theta}_{T}(\bm{x})$ to treatment differences $\bm{\delta}(\bm{x})$:}
	
	The treatment difference on outcome $k$, $\delta^{k} (\bm{x})$, is defined as the difference between the success probabilities of two treatments on outcome $k$, such that:
	\begin{flalign}\label{eq:theta2delta}
		\delta^{k} (\bm{x}) & = \theta_{1}^{k} (\bm{x}) - \theta_{0}^{k}(\bm{x}).
	\end{flalign} 
	\noindent The $K$-variate treatment difference is then $\bm{\delta}(\bm{x}) = (\delta^{1}(\bm{x}), \dots, \delta^{K}(\bm{x}))$.
	
	Multivariate treatment difference $\bm{\delta} = (\delta^{strk}, \delta^{dep})$ in the IST is a vector of the univariate treatment differences:
	\begin{flalign}\label{eq:theta2delta_IST2}
		\delta^{strk} & = \theta^{strk}_{H+A} - \theta^{strk}_{A}\\\nonumber
		\delta^{dep} & = \theta^{dep}_{H+A} - \theta^{dep}_{A}.\nonumber
	\end{flalign}
\end{enumerate}
Applying the three above-mentioned steps to each draw of the posterior sample of $\bm{\beta}$, results in a posterior sample of multivariate treatment difference $\bm{\delta} (\bm{x})$.
This sample provides estimates that can be used for prediction, where 
various measures of central tendency (e.g., a mean or high posterior density interval) can be used to summarize the sample into a point estimate.
Moreover, the sample can be used for statistical inference to generalize the conclusion to the specified (sub)population, as outlined in the next subsection.

\subsection{Posterior decision-making}\label{ss:decision}

Decisions rely on estimated treatment effects, such as differences between success probabilities, and their uncertainties.
More formally, multivariate treatment difference $\bm{\delta}$ has complete parameter spaces $\mathcal{S} \subset [-1,1]^{K}$, which is divided into a rejection region $\mathcal{S}_{R}$ and an non-rejection region $\mathcal{S}_{N}$. 
Rejection region $\mathcal{S}_{R}$ reflects the part of the parameter space that indicates the treatment difference of interest, where we would conclude that the treatments differ. 
The non-rejection region $\mathcal{S}_{N}$ refers to the part of the parameter space that would not be considered a (relevant) treatment difference. 
Rejection regions depend on the type of decision and be composed of multiple subregions if desired \citep{Ravenzwaaij2019}. 
We consider the following three (commonly used) decision types:
\begin{enumerate} 
	\item superiority with region $\mathcal{S}_{R} \in \mathcal{S}_{S}$, where the treatment is better; 
	\item inferiority with region $\mathcal{S}_{R} \in \mathcal{S}_{I}$, where the treatment is worse;
	\item two-sided with rejection region $\mathcal{S}_{R} \in \{\mathcal{S}_{S}, \mathcal{S}_{I}\}$, where the treatment can be either better or worse. 
\end{enumerate}

\begin{figure}
	\centering
	\includegraphics[width=0.9\linewidth,keepaspectratio]{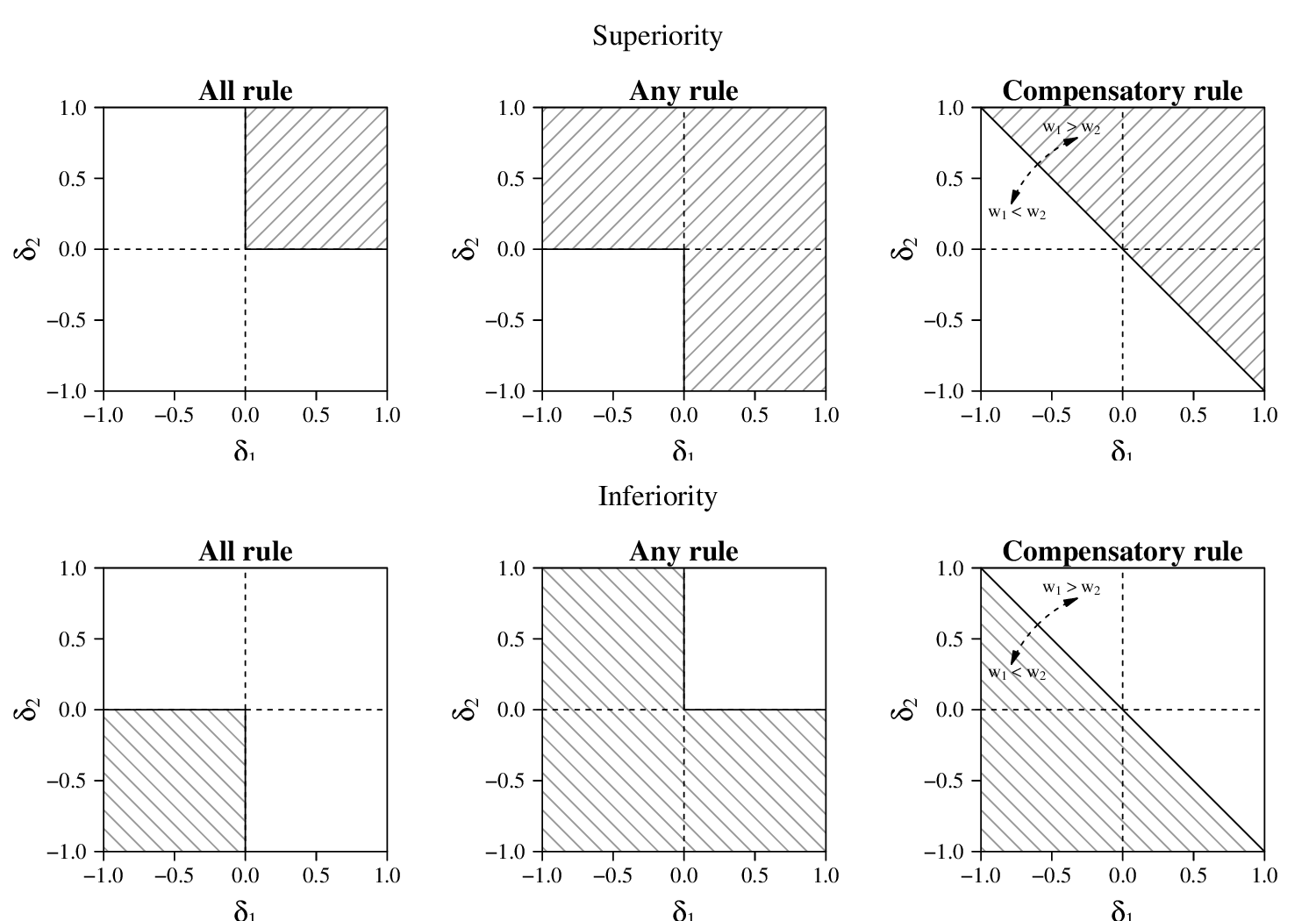}
	\caption{Bivariate superiority and inferiority spaces for the All, Any, and Compensatory ($\bm{w} = 0.50,0.50$) rules.}
	\label{fig:decision_space}
\end{figure}

We consider evidence sufficiently strong and would conclude superiority and/or inferiority when the posterior probability that treatment difference $\bm{\delta}(\bm{x})$ lies in the rejection region exceeds a prespecified decision threshold, $p_{cut}$: 
\begin{flalign}\label{eq:rejection_pop}
	p(\bm{\delta} (\bm{x}) & \in \mathcal{S}_{R} | \bm{y}) > p_{cut}.
\end{flalign}
When the functional form of the posterior distribution is unknown, the rejection probability can be concluded from an MCMC sample of $L$ draws from the posterior distribution of $\bm{\delta} (\bm{x})$.
Equation \ref{eq:rejection_pop} is then applied in practice as:
\begin{flalign}
	\frac{1}{L} \displaystyle\sum_{(l)=1}^{L} I(\bm{\delta}^{(l)} (\bm{x}) \in \mathcal{S}_{R} | \bm{y}) > p_{cut}.
\end{flalign}
In a situation with multiple dependent variables, superiority and inferiority can be defined in multiple ways, resulting in different rejection regions \citep[e.g][]{Pocock1987,Pocock1997,OBrien1984,Prentice1997,Tang1993,Zhao2007}.
Although not intended as an exhaustive overview, we list three possible rules and graphically present their rejection regions in Figure \ref{fig:decision_space}. 
Two of these rules (which we refer to as the "Any" and "All" rules) are presented as part of the regulatory guideline regarding multiple endpoints, as presented by the Food and Drug Administraction  \citet{FDA2017} and have been extensively discussed in literature \citep[e.g.,][]{ChuangStein2006,Sozu2010,Sozu2016,Xiong2005}.
The third rule ("Compensatory") is a - relatively unknown - flexible alternative that weighs benefits and risks of treatments by their (clinical) relevance \citep{Murray2016,Kavelaars2020}.
A more elaborate comparison of these rules can be found in \citet{Kavelaars2020}.

\begin{enumerate}
	
	\item \textbf{Any rule:}
	The Any rule results in superiority or inferiority when the difference between success probabilities is larger or smaller than zero respectively on at least one of the dependent variables \citep{Sozu2016}.
	The superiority and inferiority spaces are defined as:
	\begin{flalign}\label{eq:sup_any}
		\mathcal{S}^{Any}_{S}
		& =  \bm{\delta} (\bm{x}) | \max_{1<k<K} \delta^{k} (\bm{x}) > 0   \\\nonumber
		\mathcal{S}^{Any}_{I}
		& =  \bm{\delta} (\bm{x}) | \min_{1<k<K} \delta^{k} (\bm{x}) < 0.  \nonumber
	\end{flalign}
	\item \textbf{All rule:}
	The All rule results in superiority or inferiority when the difference between success probabilities is larger or smaller than zero respectively on all of the dependent variables \citep{Sozu2010}.	
	The superiority and inferiority spaces are defined as:
	\begin{flalign}\label{eq:sup_all}
		\mathcal{S}^{All}_{S} 
		& = \bm{\delta}(\bm{x}) | \min_{1<k<K} \delta^{k}(\bm{x}) > 0 \\\nonumber
		\mathcal{S}^{All}_{I} 
		& = \bm{\delta}(\bm{x}) | \max_{1<k<K} \delta^{k}(\bm{x}) < 0.  \nonumber
	\end{flalign}
	\item \textbf{Compensatory rule:}
	The Compensatory rule results in superiority or inferiority when the weighted difference between success probabilities is larger or smaller than zero respectively.
	The superiority and inferiority spaces are defined as:
	\begin{flalign}\label{eq:sup_compensatory}
		\mathcal{S}_{S}^{Comp}(\bm{w}) 
		& =  \bm{\delta}(\bm{x}) | \delta(\bm{w}, \bm{x}) > 0   \\\nonumber
		\mathcal{S}_{I}^{Comp}(\bm{w}) 
		& =  \bm{\delta}(\bm{x}) | \delta(\bm{w}, \bm{x}) < 0  \nonumber
	\end{flalign}
	\noindent where $\bm{w} = (w^{1},\dots,w^{K})$ reflect weights of $K$ treatment differences, $\delta(\bm{w}, \bm{x}) = \displaystyle\sum_{k=1}^{K} w^{k}\delta^{k} (\bm{x})$,  $ 0  \leq w^k \leq 1$ and $\displaystyle\sum_{k=1}^K w^{k}=1$ \citep{Kavelaars2020}.
\end{enumerate}

\subsection{Sample size computations}\label{ss:sample_size}

Decisions resulting from analysis with the Bayesian multivariate logistic regression framework are based on a posterior probability. %
In absence of prior information, the Bayesian posterior probability has a direct relation with the frequentist p-value: The Bayesian posterior probability equals $1 - p$ and behaves according to the well-known relationship between effect size, sample size, and decision error rates \citep{Marsman2016}. %
This allows for control of decision error rates (Type I and Type II-error) via a priori computed sample sizes. %
Methods to compute required sample sizes are available for variables that follow a multivariate Bernoulli distribution and are eligible for large sample approximation by a (multivariate) normally distributed latent variable \citep{Sozu2016,Sozu2010,Chow2017}. %
These procedures have shown to accurately control Type I rate $\alpha$ and Type II error rate $\beta$ in a Bayesian multivariate Bernoulli - Dirichlet-model on multivariate response data with a non-informative prior distribution \citep{Kavelaars2020}. %
Each of the presented decision rules in Subsection \ref{ss:decision} has an individual procedure to compute sample sizes, as discussed below. %
These equations provide insight into the required number of observations in absence of prior information and in the influence of the correlation on the sample sizes needed to obtain targeted decision error rates. %
For notational simplicity, we discard the dependence on $\bm{x}$ in the remainder of this subsection.

\subsubsection{All and Any rules}

Sample size computations for the All and Any rules were formulated in \citet{Sozu2010} and \citet{Sozu2016} respectively and rely on the assumption of a multivariate normal latent variable.
The power, $1 - \beta$, can be expressed in terms of a cumulative $K$-variate normal distribution $\bm{\Psi}_{K}$ with mean $\bm{0}$ and correlation matrix $\bm{\Sigma}$ \citep{Sozu2016}:
\begin{flalign}\label{eq:power_any_all}
	1 - \beta & = \bm{\Psi}_{K}(c^{1},\dots,c^{K}).
\end{flalign}
\noindent In Equation \ref{eq:power_any_all}, $c^{k}$ for outcome $k$ is defined by the decision rule of interest. 
Further, the off-diagonal elements of $\bm{\Sigma}$ denote (estimated) pairwise correlations between dependent variables.

\noindent For the Any rule,
\begin{flalign}\label{eq:power_any}
	c^{k} = & z_{(1 - \frac{\alpha}{K})} - \frac{(\theta^{k}_{1} - \theta^{k}_{0})}{\sqrt{\frac{
				\theta^{k}_{1} (1 - \theta^{k}_{1}) +  
				\theta^{k}_{0} (1 - \theta^{k}_{0})}{n}}}.
\end{flalign}
\noindent For the All rule, 
\begin{flalign}\label{eq:power_all}
	c^{k} = & - 
	z_{(1 - \alpha)} + 
	\frac{(\theta^{k}_{1} - \theta^{k}_{0})}{\sqrt{\frac{
				\theta^{k}_{1} (1 - \theta^{k}_{1}) +  
				\theta^{k}_{0} (1 - \theta^{k}_{0})}{n}}}.
\end{flalign}
\noindent In Equations \ref{eq:power_any} and \ref{eq:power_all}, $n$ is the sample size per treatment and $z_{(.)}$ refers to the selected $(1-\frac{\alpha}{K})$ or $(1 - \alpha)$ quantile from the univariate normal distribution.

Since the cumulative multivariate normal distribution does not have a closed-form, the sample size that satisfies targeted decision error rates can be found via the following iterative procedure proposed by Sozu et al. \citet{Sozu2010}:

\begin{enumerate}
	\item Plug in estimates of $\theta^{k}_{T}$ in Equation \ref{eq:power_any} or \ref{eq:power_all}.
	\item Plug in a starting value for $n$ in Equation \ref{eq:power_any} or \ref{eq:power_all} and calculate the power via Equation \ref{eq:power_any_all}.
	\item Repeat step 2 with gradually increasing $n$ until the power exceeds the desired level
	\item Select $n$ as the sample size per treatment group
\end{enumerate}

\subsubsection{Compensatory rule}

Sample sizes for the compensatory rule can be computed using standard methodology for large sample tests with two binomial proportions \cite[Chapter~4]{Chow2017}. 
Plugging in estimates of weighted success probabilities per treatment $T$, $\theta^{\bm{w}}_{T}$, results in:
\begin{flalign}\label{eq:power_comp}
	n & = \left[\theta_{1}^{\bm{w}} \left(1 - \theta_{1}^{\bm{w}} \right) + \theta_{0}^{\bm{w}} \left(1 - \theta_{0}^{\bm{w}}\right)\right]
	\left[\frac{z_{1-\alpha}+z_{1-\beta}}{\theta_{1}^{\bm{w}} - \theta_{0}^{\bm{w}}}\right]^2,
\end{flalign}
\noindent where $\theta^{\bm{w}}_{T} = \displaystyle\sum_{k=1}^{K} w^{k} \theta^{k}_{T}$, and $z_{1 - \beta}$ is the $(1 - \beta)$ quantile of the univariate normal distribution. 

\subsubsection{Correlation, sample size, and statistical power}\label{sss:samplesize_example}

We illustrate the relation between the sample size, the statistical power and the correlation between dependent variables with an example. %
We computed required sample sizes to obtain $80\%$ statistical power for the following bivariate ($K=2$) and trivariate ($K = 3$) outcomes, where we used different correlations ($\rho_{\theta^{k},\theta^{l}}$) and multivariate treatment differences ($\bm{\delta}$). %
Scenarios $S2$ and $S3$ had a smaller multivariate treatment difference $\bm{\delta}$ and weaker non-zero correlations than scenarios $L2$ and $L3$. %
\begin{enumerate}
	\item \bf{S2}: $K = 2$, $\rho_{\theta^{k},\theta^{l}} \in {-0.20, 0.00, 0.20}$, $\bm{\delta} = (0.20, 0.10)$
	\item \bf{S3}: $K = 3$, $\rho_{\theta^{k},\theta^{l}} \in {-0.20, 0.00, 0.20}$, $\bm{\delta} = (0.20, 0.10, 0.20)$
	\item \bf{L2}: $K = 2$, $\rho_{\theta^{k},\theta^{l}} \in {-0.40, 0.00, 0.40}$, $\bm{\delta} = (0.30, 0.20)$
	\item \bf{L3}: $K = 3$, $\rho_{\theta^{k},\theta^{l}} \in {-0.40, 0.00, 0.40}$, $\bm{\delta} = (0.30, 0.20, 0.30)$
\end{enumerate}

Table \ref{tab:SampleSize_example} shows the  the required sample sizes for these scenarios as well as the anticipated statistical power for right-sided superiority decision-making under two scenarios:
\begin{enumerate}
	\item When sample sizes computations are based on the true multivariate treatment difference $\bm{\delta}$ and the true correlation between dependent variables $\rho_{\theta^{k},\theta^{l}}$. %
	This scenario aims to highlight that statistical power can be targeted when sample size computations follow the true data generating mechanism. %
	\item When sample sizes computations are based on the true multivariate treatment difference $\bm{\delta}$ and uncorrelated dependent variables (i.e., $\rho_{\theta^{k},\theta^{l}} = 0$). %
	This scenario provides insight in anticipated error rates where the correlation is not taken into account in sample size computations. %
	This situation is equivalent to performing multiple univariate analyses on correlated dependent variables. %
\end{enumerate}
\noindent These probabilities are computed by plugging in true treatment differences and correlations, while using either the required sample size (scenario $1$) or the sample size for uncorrelated data (scenario $2$) in Equations ~\eqref{eq:power_any_all}--\eqref{eq:power_comp}).

This illustration provides five takeaways. %
First, larger effect sizes ($L2$ and $L3$) result in smaller required samples ($S2$ and $S3$) respectively. %
Second, adding an additional dependent variable has the potential to reduce sample sizes. %
Required sample sizes are larger for a three-dimensional outcome ($S3$ and $L3$) than for a two-dimensional outcome ($S2$ and $L2$). %
Third, the required sample size depends on the correlation between dependent variables. %
Compared to uncorrelated dependent variables, the Any and Compensatory rules require fewer observations when dependent variables are negatively correlated, whereas positively correlated dependent variables require more observations. %
Consequently, when sample size computations do not take non-zero correlations into account, statistical power will be larger or smaller than targeted respectively. %
Fourth, the relation between correlation and required sample size is different for different decision rules. %
Compared to the Any and Compensatory rules, the All rule shows the opposite relation between the direction of the correlation and the required sample size. %
Here, positively correlated dependent variables require a smaller number of observations than uncorrelated or negatively correlated dependent variables. %
Moreover, the All rule appears less sensitive to the correlation than the other rules. %
Sample sizes are not very different and statistical power under independence is still close to the targeted $.80$. %
Fifth, the effect of the correlation on required sample size and statistical power is larger in the scenarios where non-zero correlations are stronger ($L2$ and $L3$). %
In these scenarios, the discrepancy between the targeted power of $.80$ and the actual power is larger for non-zero correlations. %
Further, the differences between presented sample sizes for negatively correlated, uncorrelated, and positively correlated dependent variables is larger compared to the scenarios with less strong correlation ($S2$ and $S3$). %

These takeaways are in line with detailed discussions in \citet{Sozu2010,Sozu2016,FDA2017,Kavelaars2020}. %

\begin{table}[htbp]
	\small\sf\centering
	\caption{Example of required sample sizes (n) for analysis with correlated data and anticipated probabilities to conclude superiority when sample size computations use the true correlation ($p_{T}$) vs. assume uncorrelated dependent variables ($p_{U}$) under four different data-generating mechanisms (DGMs).} 
	\label{tab:SampleSize_example}
	\begin{tabular}{lrrrp{0.02cm}rrrp{0.02cm}rrr}
		\toprule
		& & \multicolumn{3}{c}{$\rho < 0$} & & \multicolumn{3}{c}{$\rho = 0$}  & \multicolumn{3}{c}{$\rho > 0$} \\
		\cmidrule(l){2-4} 
		\cmidrule(l){6-8} 
		\cmidrule(l){10-12}  
		DGM & n & $p_{T}$ & $p_{U}$& & n & $p_{T}$ & $p_{U}$& & n & $p_{T}$ & $p_{U}$\\
		\cmidrule(l){2-12} 
		& \multicolumn{11}{c}{All rule} \\
		\cmidrule(l){2-12} 
		S2    & 307 & 0.801 & 0.801 &   & 307 & 0.801 & 0.801 &   & 307 & 0.801 & 0.801 \\ 
		L2    & 77 & 0.801 & 0.801 &   & 77 & 0.803 & 0.803 &   & 76 & 0.803 & 0.808 \\ 
		S3    & 307 & 0.800 & 0.800 &   & 307 & 0.801 & 0.801 &   & 307 & 0.801 & 0.801 \\ 
		L3    & 79 & 0.801 & 0.800 &   & 79 & 0.804 & 0.804 &   & 77 & 0.803 & 0.812 \\ 
		\cmidrule(l){2-12} 
		& \multicolumn{11}{c}{Any rule} \\
		\cmidrule(l){2-12} 
		S2    & 76 & 0.801 & 0.825 &   & 81 & 0.803 & 0.803 &   & 85 & 0.802 & 0.783 \\ 
		L2    & 27 & 0.811 & 0.862 &   & 31 & 0.807 & 0.807 &   & 35 & 0.801 & 0.756 \\ 
		S3    & 51 & 0.807 & 0.850 &   & 57 & 0.804 & 0.804 &   & 64 & 0.805 & 0.760 \\ 
		L3    & 18 & 0.821 & 0.918 &   & 23 & 0.809 & 0.809 &   & 29 & 0.804 & 0.714 \\ 
		\cmidrule(l){2-12} 
		& \multicolumn{11}{c}{Compensatory rule} \\
		\cmidrule(l){2-12} 
		S2    & 53 & 0.798 & 0.845 &   & 61 & 0.801 & 0.801 &   & 68 & 0.799 & 0.760 \\ 
		L2    & 18 & 0.807 & 0.884 &   & 23 & 0.794 & 0.794 &   & 29 & 0.799 & 0.714 \\ 
		S3    & 24 & 0.796 & 0.923 &   & 37 & 0.804 & 0.804 &   & 49 & 0.802 & 0.699 \\ 
		L3    & 5 & 0.826 & 0.996 &   & 14 & 0.798 & 0.798 &   & 24 & 0.807 & 0.608 \\ 
		\bottomrule
	\end{tabular}
\end{table}

\section{Estimating conditional average treatment effects}\label{s:heterogeneity}

In the proposed framework, treatment heterogeneity can be captured by joint response probabilities that reflect conditional average treatment effects and thus depend on prespecified characteristics of a subpopulation of interest. 
We describe two ways to represent subpopulations: by fixed covariate values or by a prespecified interval of the covariate distribution(s).
Both representations have their own applications.
Fixed values of covariates may be relevant when we wish to investigate treatment effects based on individual patients or on patient populations that can be accurately represented by a single number of the covariate (such as a mean or a level of a discrete variable).
Intervals of covariate distributions may be sensible in particular when multiple consecutive covariate values are sufficiently exchangeable to estimate a marginal treatment effect among a population specified by this range. 
Although such intervals can be specified for discrete covariates as well, their use is particularly reasonable with continuous covariates, as intervals are inherently consistent with the idea of continuity.

We will discuss procedures to estimate conditional average treatment effects based on fixed values and based on intervals in more detail in the remainder of this subsection. 
In these discussions, we use a linear predictor $\bm{\psi}^{q}_{i} (\bm{x})$ (cf. Equation \ref{eq:psi}) that distinguishes between treatments via a treatment indicator and allows for interaction between the treatment and a covariate. 
For such a model that includes a single population characteristic $x$, $\bm{x} = (z, T, zT)$ and $\psi^{q}_{T} (\bm{x})$ is defined as:  
\begin{flalign}\label{eq:psi_spec}
	\psi^{q}_{T} (\bm{x}) = &\beta^{q}_{0} + \beta^{q}_{1} T + \beta^{q}_{2} z + \beta^{q}_{3} zT.
\end{flalign} 
\subsection{Fixed values of covariate}\label{proc:values}

For a patient population with fixed values of patient covariates, a posterior sample of joint response probabilities $\bm{\phi}_{T} (\bm{x})$ can be found by plugging in a vector of fixed covariate values $\bm{x}$ in linear predictor $\bm{\psi}^{(l)}_{T} (\bm{x})$.
Subsequently applying the multinomial logistic link function in Equation \ref{eq:psi2phi} to each $\bm{\psi}^{(l)}_{T} (\bm{x})$ results in joint response probability $\bm{\phi}^{(l)}_{T} (\bm{x})$ for treatment $T$.
Applying these steps each posterior draw $(l)$ of regression coefficients $\bm{\beta}^{(l)}$ results in a sample of posterior joint response probabilities.
The procedure is presented in Algorithm \ref{alg:procedure_fixed_values} in Appendix \ref{app:alg_transformation}.

\subsection{Marginalization over a distribution of covariates}\label{proc:analytical}
When the population is characterized by a range of covariates, the treatment effect can be marginalized over the interval under consideration, based on available information regarding the distribution of the covariate.

A sample of covariate data can be used as input for marginalization.
Empirical marginalization involves repeating the fixed values procedure for each subject in the sample to obtain a sample of joint response probabilities for each posterior draw $(l)$. 
Averaging the resulting sample of joint response probabilities per treatment results in a marginal joint response probability $\bm{\phi}^{(l)}_{T} (\bm{x})$ for draw $(l)$.
The procedure is presented in Algorithm \ref{alg:procedure_empirical} in the online supplemental materials.
Empirical marginalization is computationally efficient for patient populations defined by intervals of more than one continuous covariate.
Note however that the procedure is prone to sampling variability in $\bm{x}$ and that estimation might depend on the availability of cases with the selected covariate values. 
Increasing the specificity of subpopulations -  often resulting from a higher number of included covariates and/or a limited interval size - will reduce the number of available observations eligible for inclusion\footnote{If this is the case, (numerical) integration can be an alternative to interpolate the conditional treatment effect distribution of interest.}.


\section{Numerical evaluation}\label{s:evaluation}

The current section presents an evaluation of the performance of the proposed multivariate logistic regression procedure. 
The goal of the evaluation was twofold and we aimed to demonstrate:
\begin{enumerate}
	\item how well the obtained regression coefficients and treatment effects correspond to their true values to examine bias; 
	\item how often the decision procedure results in an (in)correct superiority conclusion to learn about decision error rates when sample sizes are estimated a priori.
\end{enumerate}

\subsection{Setup}

\subsubsection{Conditions}

The performance of the framework was evaluated in a treatment comparison based on one covariate and two dependent variables. 
In Appendix \ref{app:evaluation_k3}, we present an evaluation of the performance with three dependent variables.
Six aspects were varied: the analysis procedure, the effect size, measurement level of the covariate, the correlation between dependent variables, the (sub)population, and the decision rule.
Each of these factors will be discussed in the following paragraphs.

\paragraph{Analysis procedure}
\noindent We present three Bayesian analysis procedures:
\begin{enumerate}
	\item \textbf{Multivariate logistic regression analysis (mLR)}: 
	We analyzed the generated data via the proposed Bayesian multivariate logistic regression model presented in Section \ref{s:model}. 
\end{enumerate}
The performance of the mLR-model was compared to two reference approaches:
\begin{enumerate}
	\setcounter{enumi}{2}
	\item \textbf{Multivariate Bernoulli analysis (mB)}: 	
	To demonstrate the gain of a multivariate regression approach over multivariate subgroup analysis (i.e., multivariate stratified analysis), we fitted the unconditional Bayesian multivariate Bernoulli model in \citet{Kavelaars2020} to the data as well. 
	Whereas the multivariate Bernoulli model takes the correlation between dependent variables into account, the multivariate Bernoulli model computes conditional average treatment effects via stratified multivariate analysis: the multivariate Bernoulli model only uses the response data from observations that belong to the (sub)population of interest. 
	Hence, the estimation of ATEs uses the full sample of response data, whereas CATEs are estimated based on a subsample of response data.
	Samples of treatment-specific joint response probabilities $\bm{\phi}_{T}$ could be drawn directly from a posterior Dirichlet distribution with parameters $\bm{\alpha}_{T}^{n} = \bm{\alpha}^{0} + \{\displaystyle\sum_{i=1}^{n} I(T_{i} = T) I(\bm{y}_{i} =\bm{H}_{q\cdot})\}_{q = 1}^{Q}$, where $\bm{\alpha}^{0}$ is a vector of $Q$ prior hyperparameters.
	
	\item \textbf{Univariate logistic regression (uLR)}: 
	To demonstrate the added value of a multivariate model over multiple univariate models, we fitted Bayesian univariate logistic regression models from \citet{polson2013} to the individual dependent variables for the scenario with two dependent variables.
	This univariate model is a special case of the multivariate model presented in Section \ref{s:model} and Appendix \ref{app:posterior_computation}.
	While these regression-based models use the full sample of data to estimate conditional average treatment effects among subpopulations, they cannot capture correlations between dependent variables.
	
\end{enumerate}

\paragraph{Datagenerating mechanisms: effect size, measurement level of covariate, and correlation}
{
	We included treatment differences of four different sizes that varied in heterogeneity: 
	
	\begin{enumerate}
		\item 
		\textbf{Effect size $\bm{1.1}$ \& $\bm{1.2}$:} A homogeneous treatment effect, with average and conditional treatment differences of zero. 
		This scenario aims to demonstrate the Type I error rate under a least favorable treatment difference for the Any and Compensatory rules
		in the trial as well as the subpopulation.
		
		\item 
		\textbf{Effect size $\bm{2.1}$ \& $\bm{2.2}$:} A heterogeneous treatment effect, with an average treatment difference of zero and a conditional treatment effect larger than zero.
		
		\item 
		\textbf{Effect size $\bm{3.1}$ \& $\bm{3.2}$:} A heterogeneous treatment treatment effect, with one average and both conditional treatment differences larger than zero.
		The conditional treatment difference is larger than the average treatment difference. 
		The effect size is chosen to compare power of different methods, when the sample size should not lead to underpowerment for any of the approaches to the estimation of conditional average treatment effects. 
		The effect size of the conditional average treatment effect reflects the least favorable average treatment effect for a right-sided test of the All rule and should result in a Type I error rate equal to the chosen level of $\alpha$ if the sample size is sufficiently large.
		
		\item 
		\textbf{Effect size $\bm{4.1}$ \& $\bm{4.2}$:}A heterogeneous treatment treatment effect, with one average and both conditional treatment differences larger than zero.
		The conditional treatment difference is smaller than the average treatment effect. 
		The effect size is chosen such that the expected sample size after stratification of the study sample is smaller than the required sample for evaluation of the conditional treatment effect and aims to investigate the statistical power of regression-based methods when stratification leads to underpowered decisions.
		Similar to effect size $3.1$/$3.2$, the effect size of the conditional average treatment effect reflects the least favorable effect for a right-sided test of the All rule and should result in a Type I error rate equal to the chosen level of $\alpha$ if the sample size is sufficiently large.
		
	\end{enumerate} 
}\label{response:DGMs}

For each of these four effect sizes, we varied the measurement level of the covariate and created a model with a binary covariate and a model with a continuous covariate. 
Further, we specified three pairwise correlations for the dependent variables: a negative correlation ($\rho_{\theta^{k},\theta^{l}} = -.20$), no correlation ($\rho_{\theta^{k},\theta^{l}} = .00$), and a positive correlation ($\rho_{\theta^{k},\theta^{l}} = .20$). 
These pairwise correlations were identical for all dependent variable pairs and were specified for the conditional average treatment effects ($x = 0$ and $x = 1$ for the dichotomous covariate; at $x = -1$ and $x = 1$ for the continuous covariate).
The correlation structures and effect sizes of the conditional average treatment effects determine, together with the probability distribution of the covariates, the correlation and effect size of average treatment effects.

These four effect sizes, two measurement levels of the covariate, and three correlation structures resulted in the $4 \times 2 \times 3 = 24$ data generating mechanisms (DGMs) presented in Table \ref{tab:Delta.k2}.

\begin{table}[htbp]
	\small\sf\centering
	\caption{Parameters of average treatment effects (ATEs) in the trial and conditional average treatment effects (CATEs) in a subpopulation for two outcome variables. } 
	\label{tab:Delta.k2}
	\begin{tabular}{llp{0.02cm}rrrp{0.02cm}rrr}
		\toprule
		
		&  & & \multicolumn{3}{c}{ATE} & & \multicolumn{3}{c}{CATE} \\ 
		\cmidrule(lr){3-6} 
		\cmidrule(l){7-10} 
		ES &  &  & \multicolumn{1}{l}{$(\delta_{1}, \delta_{2})$} & \multicolumn{1}{l}{$\delta (\bm{\mathsf{w}})$} & \multicolumn{1}{l}{$\rho_{\theta^{k},\theta^{l}}$} &  & \multicolumn{1}{l}{$(\delta_{1}, \delta_{2})$} & \multicolumn{1}{l}{$\delta (\bm{\mathsf{w}})$} & \multicolumn{1}{l}{$\rho_{\theta^{k},\theta^{l}}$} \\
		\midrule
		1.1 & D &   & (0.000,  0.000) &  0.000 & -0.160 &   & (0.000,  0.000) &  0.000 & -0.200 \\ 
		&  &   &   &   &  0.030 &   &   &   &  0.000 \\ 
		&  &   &   &   &  0.220 &   &   &   &  0.200 \\ 
		1.2 & C &   & (0.000,  0.000) &  0.000 & -0.163 &   & (0.000,  0.000) &  0.000 & -0.207 \\ 
		&  &   &   &   &  0.028 &   &   &   &  0.002 \\ 
		&  &   &   &   &  0.219 &   &   &   &  0.208 \\ 
		2.1 & D &   & (0.000,  0.000) &  0.000 & -0.154 &   & (0.250,  0.150) &  0.225 & -0.200 \\ 
		&  &   &   &   &  0.037 &   &   &   &  0.000 \\ 
		&  &   &   &   &  0.229 &   &   &   &  0.200 \\ 
		2.2 & C &   & (0.000,  0.000) &  0.000 & -0.157 &   & (0.116,  0.069) &  0.104 & -0.206 \\ 
		&  &   &   &   &  0.036 &   &   &   &  0.003 \\ 
		&  &   &   &   &  0.228 &   &   &   &  0.207 \\ 
		3.1 & D &   & (0.100,  0.000) &  0.075 & -0.152 &   & (0.300,  0.200) &  0.275 & -0.200 \\ 
		&  &   &   &   &  0.040 &   &   &   &  0.000 \\ 
		&  &   &   &   &  0.232 &   &   &   &  0.200 \\ 
		3.2 & C &   & (0.100,  0.000) &  0.075 & -0.155 &   & (0.196,  0.093) &  0.170 & -0.205 \\ 
		&  &   &   &   &  0.038 &   &   &   &  0.003 \\ 
		&  &   &   &   &  0.231 &   &   &   &  0.206 \\ 
		4.1 & D &   & (0.350,  0.000) &  0.263 & -0.197 &   & (0.200,  0.000) &  0.150 & -0.200 \\ 
		&  &   &   &   &  0.000 &   &   &   &  0.000 \\ 
		&  &   &   &   &  0.197 &   &   &   &  0.200 \\ 
		4.2 & C &   & (0.350,  0.000) &  0.263 & -0.197 &   & (0.288,  0.000) &  0.216 & -0.202 \\ 
		&  &   &   &   &  0.000 &   &   &   &  0.000 \\ 
		&  &   &   &   &  0.197 &   &   &   &  0.202 \\ 
		\midrule 
		\multicolumn{10}{l}{Es = Effect size, D = Discrete covariate, C = Continuous covariate} \\
		\bottomrule
	\end{tabular}
\end{table}

\paragraph{Treatment effects and (sub)populations}
We estimated three different treatment effects:
\begin{enumerate}
	\item An average treatment effect (ATE) among the trial population. 
	The trial population with a discrete covariate was defined by a binomially distributed covariate with a probability of $0.50$.
	The trial population with a continuous covariate was defined by a covariate that followed a standard normal distribution. 
	
	\item A conditional average treatment effect (CATE) among a subpopulation defined by a sample of covariate an interval of a continuous covariate. 
	This treatment effect was also estimated among patients scoring low on the covariate, but this time the subpopulation was defined as all values between the mean and one standard deviation below the mean. 
	Note that the discrete covariate could not be assigned an interval, since subsetting a binary variable inherently results in a single value. 
	
	\item A conditional average treatment effect (CATE) among a subpopulation defined by a fixed value of a covariate. 
	The treatment effect was estimated among patients scoring low on the covariate and was described by a value of $0$ (discrete covariate) or $-1$ (continuous covariate).
	
\end{enumerate}

\paragraph{Decision rules and sample size}

We applied the three decision rules from Subsection \ref{sss:decision_rules}:
\begin{enumerate}
	\item \textbf{Any rule} 
	\item \textbf{All rule}
	\item \textbf{Compensatory rule} with unequal weights ($\bm{w} = (0.75,0.25)$)
\end{enumerate}

We computed sample sizes per treatment group via the procedures from Subsection \ref{ss:sample_size} for conditions with non-zero true average treatment effects targeting at a power of $.80$ and a right-sided $\alpha$ of $.05$.
If the true average treatment difference was equal to zero, we used $n=1,000$ per treatment group.
The sample size for the average treatment effect was thus leading for the analysis of both average and conditional average treatments effects.
As a result, the power of conditional treatment effects was not targeted at $.80$, but should exceed this target when the required sample size for a CATE was larger than the sample size for an ATE.
Similarly, the power of CATEs with a sample size smaller than the ATE sample size should be lower than $.80$.
The required sample sizes are presented in Table \ref{tab:SampleSizes.k2}.
In these tables, we also included a) the required sample size for the conditional average treatment effect in the subpopulation; and b) the sample size after stratification of the trial population. 
The sample size after stratification is the expected size in subpopulation analysis of a) response data in a stratified analysis approach; and b) covariate data in empirical marginalization. 

\begin{table}[htbp]
	\small\sf\centering
	\caption{Required sample sizes to evaluate the average treatment effect (ATE) and conditional treatment effect (CATE) for two outcome variables.} 
	\label{tab:SampleSizes.k2}
	\begin{tabular}{llp{0.02cm}rrrp{0.02cm}rrrp{0.02cm}rrr}
		\toprule
		&  & & \multicolumn{3}{c}{All} & & \multicolumn{3}{c}{Any} & & \multicolumn{3}{c}{Compensatory} \\
		
		\cmidrule(lr){4-6}
		\cmidrule(l){8-10}
		\cmidrule(l){12-14} ES & $\rho_{k^{\theta},l^{\theta}}$  & & ATE & CATE & Sub & & ATE & CATE & Sub & & ATE & CATE & Sub \\
		\midrule
		1.1 & $<0$ &   & - & - & 500 &   & - & - & 500 &   & - & - & 500 \\ 
		& $\approx 0$ &   & - & - & 500 &   & - & - & 500 &   & - & - & 500 \\ 
		& $>0$ &   & - & - & 500 &   & - & - & 500 &   & - & - & 500 \\ 
		1.2 & $<0$ &   & - & - & 342 &   & - & - & 342 &   & - & - & 342 \\ 
		& $\approx 0$ &   & - & - & 342 &   & - & - & 342 &   & - & - & 342 \\ 
		& $>0$ &   & - & - & 342 &   & - & - & 342 &   & - & - & 342 \\ 
		2.1 & $<0$ &   & - & 136 & 500 &   & - & 45 & 500 &   & - & 32 & 500 \\ 
		& $\approx 0$ &   & - & 136 & 500 &   & - & 48 & 500 &   & - & 36 & 500 \\ 
		& $>0$ &   & - & 136 & 500 &   & - & 51 & 500 &   & - & 40 & 500 \\ 
		2.2 & $<0$ &   & - & 658 & $\textbf{342}$ &   & - & 215 & 342 &   & - & 154 & 342 \\ 
		& $\approx 0$ &   & - & 649 & $\textbf{342}$ &   & - & 229 & 342 &   & - & 175 & 342 \\ 
		& $>0$ &   & - & 644 & $\textbf{342}$ &   & - & 245 & 342 &   & - & 196 & 342 \\ 
		3.1 & $<0$ &   & - & 77 & 500 &   & 381 & 29 & 191 &   & 309 & 21 & 155 \\ 
		& $\approx 0$ &   & - & 77 & 500 &   & 385 & 31 & 193 &   & 349 & 23 & 175 \\ 
		& $>0$ &   & - & 76 & 500 &   & 387 & 33 & 194 &   & 388 & 26 & 194 \\ 
		3.2 & $<0$ &   & - & 358 & $\textbf{342}$ &   & 379 & 81 & 130 &   & 307 & 56 & 105 \\ 
		& $\approx 0$ &   & - & 358 & $\textbf{342}$ &   & 383 & 86 & 131 &   & 347 & 65 & 119 \\ 
		& $>0$ &   & - & 356 & $\textbf{342}$ &   & 386 & 91 & 132 &   & 386 & 73 & 132 \\ 
		4.1 & $<0$ &   & - & - & 500 &   & 28 & 93 & $\textbf{14}$ &   & 22 & 73 & $\textbf{11}$ \\ 
		& $\approx 0$ &   & - & - & 500 &   & 28 & 93 & $\textbf{14}$ &   & 25 & 83 & $\textbf{13}$ \\ 
		& $>0$ &   & - & - & 500 &   & 28 & 94 & $\textbf{14}$ &   & 28 & 93 & $\textbf{14}$ \\ 
		4.2 & $<0$ &   & - & - & 342 &   & 28 & 43 & $\textbf{10}$ &   & 22 & 34 & $\textbf{8}$ \\ 
		& $\approx 0$ &   & - & - & 342 &   & 28 & 44 & $\textbf{10}$ &   & 25 & 39 & $\textbf{9}$ \\ 
		& $>0$ &   & - & - & 342 &   & 28 & 44 & $\textbf{10}$ &   & 28 & 43 & $\textbf{10}$ \\ 
		\midrule 
		
		\multicolumn{14}{l}{Sub = expected size of subsample} \\  
		\multicolumn{14}{l}{Bold-faced subsamples are smaller than required for estimation of the CATE} \\ \bottomrule
	\end{tabular}
\end{table}

\subsubsection{Procedure}

\paragraph{Data generation}

For each data generating mechanism and each unique (decision-rule specific) sample size, we sampled $1000$ datasets.
We generated one covariate $x$ and included an interaction between the treatment and the covariate as well, resulting in the following linear predictor $\psi^{q}_{i}$:
\begin{flalign}
	\psi^{q}_{i} (x_{i}) & = \beta^{q}_{0} + \beta^{q}_{T} T_{i} + \beta^{q}_{1} z_{i} + \beta^{q}_{2} z_{i} T_{i}.
\end{flalign} 
\noindent To generate response data, we first applied the multinomial logistic link function (Equation \ref{eq:psi2phi}) to each true linear predictor $\bm{\psi}_{i} (x_{i})$ to obtain joint response probabilities $\bm{\phi}_{i} (x_{i})$ for each subject $i$.
Next, we sampled response vector $\bm{y}_{i}|\bm{x}_{i}$ from a multinomial distribution with probabilities $\bm{\phi}_{i} (x_{i})$. 

\paragraph{Prior distribution}\label{sss:evaluation_prior}

{
	We specified diffuse prior distributions. %
	This is motivated by the idea that obtained the posterior distributions are then completely based on the information in the data, which is a common choice in default Bayesian analyses. %
	For the multivariate logistic regression analysis, we set multivariate normally distributed prior with means $\bm{b}^{q} = \bm{0}$ and variance matrix $\bm{B}^{0q} = \text{diag }(1e^{1}, \dots, 1e^{1})$ for all regression coefficients. %
	Prior covariances between regression coefficients were set at zero, implying that regression coefficients were independent a priori. %
	For the univariate logistic regression analysis we used univariate normally distributed priors with a mean of $0$ and a variance of $1e^{1}$ for all parameters. %
	The specified variance parameters of regression coefficients were motivated by work of \citet{Gelman2008}. %
	These authors recommend to choose a variance parameter that results in realistic support for the probability parameter after non-linear transformations in logistic regression and has sufficient information to stabilize posterior computations. %
	For a reference approach, we used a Dirichlet prior distribution with hyperparameters $\bm{\alpha}^{0} = \bm{0.01}$. %
	This prior is close to the improper Haldane prior ($\bm{\alpha} = \bm{0}$), which is considered the least informative prior distribution for bi- or multinomially distributed data, results in a posterior mean equal to the maximum likelihood estimator, and corresponds to a uniform prior on the log-odds scale \citep{Tuyl2008,Kerman2011}. %
	The small deviation from the Haldane prior makes the prior distribution proper and ensures that cell probabilities can be sampled from the Dirichlet distribution when cells have no observations \citep{Kavelaars2020}. %
}

\paragraph{Gibbs sampling}

The regression coefficients in response categories $1,\dots,(Q-1)$ were estimated via the Gibbs sampler detailed in the 
online supplemental materials. 
We ran two MCMC-chains with $L = 10,000$ iterations plus $1,000$ burnin iterations. 
{
	We visually inspected traceplots of MCMC-chains and used multivariate Gelman-Rubin convergence diagnostics to assess convergence \citep{Gelman1992,Brooks1998}. %
	As these traceplots showed satisfactory overlap between chains and the convergence diagnostics were all between $1.00$ and $1.10$, we concluded that there were no issues with convergence. %
}\label{response:convergence_sim}

\paragraph{Transformation and decision-making}\label{sss:decision_rules}

We applied the procedures from Subsections \ref{ss:transformation} and \ref{ss:decision} to arrive at a decision. 
In marginalization, we included the selection of subjects that belonged to the subpopulation.
We performed a right-sided (superiority) test aiming at a Type I-error rate of $\alpha = .05$.
We used a decision threshold $p_{cut} = 1 - \alpha = 0.95$ (Compensatory and All rules) and a for multiple tests corrected $p_{cut} = 1 - \frac{\alpha}{K} = 0.975$ (Any rule) \citep{Marsman2016, Kavelaars2020, Sozu2016}.

\subsubsection{Software}\label{sss:software}
We conducted our analyses in \texttt{R} \citep{RCT2020}. 
We drew variables from the multivariate normal, Pólya-Gamma, and Dirichlet distributions with the \texttt{MASS}, \texttt{pgdraw} and \texttt{MCMCpack} packages respectively \citep{Venables2002, Makalic2016, Martin2011}. 
We used the \texttt{coda} package to explore MCMC chains \citep{Plummer2006}.
The simulation procedure was parallellized using the foreach and doParallel packages \citep{Microsoft2020, Microsoft2020a}. 
\LaTeX tables were created with the \texttt{xtable} package \citep{Dahl2019}.

\subsection{Results}

\subsubsection{Bias}\label{s:results_bias}
Bias of multivariate and weighted treatment differences was negligible ($< |.01|$) in most conditions, implying that Bayesian multivariate logistic regression analysis was generally able to reproduce true treatment effects.
However, the estimation of average treatment effects under effect sizes $4.1$ and $4.2$ resulted in slightly biased treatment differences for the Any and Compensatory rules.
As shown in Table \ref{tab:Bias.k2}, these absolute biases ranged up to $|0.04|$.
These biases were produced in both univariate and multivariate logistic regression analysis, but not in multivariate Bernoulli analysis. 
Conditional average treatment effects were estimated with comparable patterns of bias and a maximum of $|.025|$.
This bias showed up in conditions with a small sample, which is a well-documented property of logistic regression in general \citep{Nemes2009}.

This bias in treatment differences could be traced back to bias in regression coefficients.
Mean estimates of regression coefficients were asymptotically unbiased, implying that bias was negligible ($<|0.01|$) in conditions with a sufficiently large sample.
We observed some bias in conditions with smaller samples (ES $3.1$, $3.2$, $4.1$, and $4.2$ under the Any and Compensatory decision rules).
We can conclude that bias in regression coefficients was not necessarily problematic for our actual parameters of interest, namely success probabilities and differences between them. 
Even when regression coefficients had a small bias ($< |0.20|$ on the log-odds scale), success probabilities and treatment differences could be estimated without bias ($<|0.01|$), similar to the conditions without biased regression coefficients.
This was the case for ES $3.1$ and $3.2$ under sample sizes of the Any and Compensatory rules.
Only more severe bias of regression coefficients ($< |0.57|$ on the log-odds scale) in conditions with smaller sample sizes was not fully corrected in the transformation steps.  
This was seen in ES $4.1$ and $4.2$ under sample sizes of the Any and Compensatory rules.

\begin{table}[htbp]
	\small\sf\centering
	\caption{Bias in average treatment differences of effect size (ES) $4.1$ and $4.2$ by decision rule.} 
	\label{tab:Bias.k2}
	\begin{tabular}{lcccc}
		\toprule  
		& & \multicolumn{3}{c}{All rule} \\ 
		\cmidrule(l){3-5} & & \multicolumn{1}{c}{uLR} & 
		\multicolumn{1}{c}{mB} & \multicolumn{1}{c}{mLR} \\
		ES & $\rho_{\theta^{k},\theta^{l}}$ & \multicolumn{1}{c}{($\delta^{1},\delta^{2}$)} & \multicolumn{1}{c}{($\delta^{1},\delta^{2}$)} & \multicolumn{1}{c}{($\delta^{1},\delta^{2}$)}\\
		\cmidrule(l){2-5} 
		
		4.1 & $<0$ & (\ 0.000,\ \:0.000) & (-0.002,\ \:0.001) & (\ 0.000,\:-0.002) \\ 
		& $\approx 0$ & (\ 0.000,\ \:0.000) & (\ 0.000,\ \:0.000) & (-0.001,\:-0.001) \\ 
		& $>0$ & (-0.001,\ \:0.000) & (-0.001,\:-0.001) & (-0.001,\ \:0.000) \\ 
		4.2 & $<0$ & (-0.002,\ \:0.000) & (\ 0.002,\:-0.001) & (\ 0.000,\:-0.002) \\ 
		& $\approx 0$ & (-0.001,\:-0.001) & (-0.001,\ \:0.001) & (-0.002,\:-0.001) \\ 
		& $>0$ & (-0.001,\:-0.001) & (\ 0.001,\:-0.001) & (-0.001,\ \:0.001) \\ 
		
		\midrule
		& & \multicolumn{3}{c}{Any rule} \\ 
		\cmidrule(l){3-5}
		& & \multicolumn{1}{c}{uLR} & 
		\multicolumn{1}{c}{mB} & \multicolumn{1}{c}{mLR} \\
		ES & $\rho_{\theta^{k},\theta^{l}}$ & \multicolumn{1}{c}{($\delta^{1},\delta^{2}$)} & \multicolumn{1}{c}{($\delta^{1},\delta^{2}$)} & \multicolumn{1}{c}{($\delta^{1},\delta^{2}$)}\\
		\cmidrule(l){1-5} 
		4.1 & $<0$ & (-0.013,\ \:0.001) & (-0.001,\ \:0.005) & (-0.024,\:-0.011) \\ 
		& $\approx 0$ & (-0.009,\:-0.002) & (\ 0.001,\ \:0.001) & (-0.023,\:-0.016) \\ 
		& $>0$ & (-0.006,\ \:0.001) & (\ 0.002,\ \:0.004) & (-0.028,\:-0.007) \\ 
		4.2 & $<0$ & (-0.018,\:-0.009) & (-0.005,\ \:0.003) & (-0.031,\:-0.019) \\ 
		& $\approx 0$ & (-0.014,\ \:0.003) & (-0.002,\:-0.001) & (-0.032,\:-0.011) \\ 
		& $>0$ & (-0.018,\:-0.002) & (-0.001,\ \:0.005) & (-0.030,\:-0.008) \\ 
		\midrule
		& & \multicolumn{3}{c}{Compensatory rule} \\ 
		\cmidrule(l){3-5} 
		& & \multicolumn{1}{c}{uLR} & 
		\multicolumn{1}{c}{mB} & \multicolumn{1}{c}{mLR} \\
		ES & $\rho_{\theta^{k},\theta^{l}}$ & \multicolumn{1}{c}{$\delta (\textbf{w})$} & \multicolumn{1}{c}{$\delta (\textbf{w})$} & \multicolumn{1}{c}{$\delta (\textbf{w})$}\\
		\cmidrule(l){1-5} 
		4.1 & $<0$ & \multicolumn{1}{c}{-0.006} & \multicolumn{1}{c}{\ 0.000} & \multicolumn{1}{c}{-0.030} \\ 
		& $\approx 0$ & \multicolumn{1}{c}{-0.012} & \multicolumn{1}{c}{\ 0.000} & \multicolumn{1}{c}{-0.019} \\ 
		& $>0$ & \multicolumn{1}{c}{-0.004} & \multicolumn{1}{c}{-0.006} & \multicolumn{1}{c}{-0.015} \\ 
		4.2 & $<0$ & \multicolumn{1}{c}{-0.018} & \multicolumn{1}{c}{-0.003} & \multicolumn{1}{c}{-0.040} \\ 
		& $\approx 0$ & \multicolumn{1}{c}{-0.017} & \multicolumn{1}{c}{-0.003} & \multicolumn{1}{c}{-0.029} \\ 
		& $>0$ & \multicolumn{1}{c}{-0.013} & \multicolumn{1}{c}{\ 0.003} & \multicolumn{1}{c}{-0.024} \\ 
		\midrule 
		\multicolumn{5}{l}{uLR = univariate logistic regression} \\
		\multicolumn{5}{l}{mB = multivariate Bernoulli} \\
		\multicolumn{5}{l}{mLR = multivariate logistic regression} \\
		\bottomrule 
	\end{tabular}
\end{table}

\subsubsection{Decision error rates}\label{s:results_decisionerrors}

\begin{table}[htbp]
	\small\sf\centering
	\caption{Proportions of superiority decisions for ATEs with two outcome variables by data-generating mechanism, correlation, and decision rule.} 
	\label{tab:pReject.ATE.k2}
	\begin{tabular}{lp{0.02cm}rrrp{0.02cm}rrrp{0.02cm}rrr}
		\toprule 
		& & \multicolumn{3}{c}{$\rho < 0$} & & \multicolumn{3}{c}{$\rho = 0$} & & \multicolumn{3}{c}{$\rho > 0$} \\
		\cmidrule(l){3-5} 
		\cmidrule(l){7-9} 
		\cmidrule(l){11-13} 
		ES  & & uLR & mB & mLR & & uLR & mB & mLR & & uLR & mB & mLR \\
		\cmidrule(l){3-13} 
		\multicolumn{13}{c}{Rule = All} \\
		\cmidrule(l){3-13} 
		1.1 &   & 0.000 & 0.004 & 0.000 &   & 0.000 & 0.005 & 0.001 &   & 0.004 & 0.005 & 0.007 \\ 
		1.2 &   & 0.003 & 0.002 & 0.001 &   & 0.000 & 0.005 & 0.002 &   & 0.006 & 0.006 & 0.005 \\ 
		2.1 &   & 0.000 & 0.001 & 0.003 &   & 0.002 & 0.004 & 0.004 &   & 0.006 & 0.005 & 0.008 \\ 
		2.2 &   & 0.002 & 0.003 & 0.000 &   & 0.007 & 0.002 & 0.005 &   & 0.003 & 0.005 & 0.010 \\ 
		3.1 &   & 0.064 & 0.051 & 0.046 &   & 0.066 & 0.046 & 0.056 &   & 0.054 & 0.043 & 0.046 \\ 
		3.2 &   & 0.051 & 0.048 & 0.055 &   & 0.050 & 0.057 & 0.050 &   & 0.049 & 0.052 & 0.061 \\ 
		4.1 &   & 0.059 & 0.051 & 0.042 &   & 0.059 & 0.044 & 0.044 &   & 0.052 & 0.046 & 0.053 \\ 
		4.2 &   & 0.051 & 0.058 & 0.045 &   & 0.045 & 0.041 & 0.051 &   & 0.044 & 0.049 & 0.053 \\ 
		\cmidrule(l){3-13} 
		\multicolumn{13}{c}{Rule = Any} \\
		\cmidrule(l){3-13} 
		1.1 &   & 0.052 & 0.046 & 0.054 &   & 0.060 & 0.064 & 0.060 &   & 0.059 & 0.047 & 0.050 \\ 
		1.2 &   & 0.054 & 0.055 & 0.043 &   & 0.035 & 0.042 & 0.050 &   & 0.038 & 0.053 & 0.049 \\ 
		2.1 &   & 0.063 & 0.053 & 0.059 &   & 0.055 & 0.044 & 0.045 &   & 0.052 & 0.049 & 0.049 \\ 
		2.2 &   & 0.059 & 0.055 & 0.062 &   & 0.059 & 0.045 & 0.062 &   & 0.046 & 0.048 & 0.060 \\ 
		3.1 &   & $\textbf{0.807}$ & $\textbf{0.802}$ & $\textbf{0.789}$ &   & $\textbf{0.810}$ & $\textbf{0.812}$ & $\textbf{0.806}$ &   & $\textbf{0.796}$ & $\textbf{0.787}$ & $\textbf{0.791}$ \\ 
		3.2 &   & $\textbf{0.814}$ & $\textbf{0.790}$ & $\textbf{0.807}$ &   & $\textbf{0.819}$ & $\textbf{0.811}$ & $\textbf{0.791}$ &   & $\textbf{0.811}$ & $\textbf{0.803}$ & $\textbf{0.815}$ \\ 
		4.1 &   & $\textbf{0.804}$ & $\textbf{0.756}$ & $\textbf{0.781}$ &   & $\textbf{0.816}$ & $\textbf{0.775}$ & $\textbf{0.787}$ &   & $\textbf{0.808}$ & $\textbf{0.780}$ & $\textbf{0.777}$ \\ 
		4.2 &   & $\textbf{0.790}$ & $\textbf{0.749}$ & $\textbf{0.793}$ &   & $\textbf{0.806}$ & $\textbf{0.774}$ & $\textbf{0.770}$ &   & $\textbf{0.781}$ & $\textbf{0.754}$ & $\textbf{0.785}$ \\ 
		\cmidrule(l){3-13} 
		\multicolumn{13}{c}{Rule = Compensatory} \\
		\cmidrule(l){3-13} 
		1.1 &   & 0.049 & 0.056 & 0.054 &   & 0.059 & 0.069 & 0.050 &   & 0.076 & 0.047 & 0.048 \\ 
		1.2 &   & 0.045 & 0.041 & 0.056 &   & 0.045 & 0.040 & 0.051 &   & 0.063 & 0.047 & 0.055 \\ 
		2.1 &   & 0.053 & 0.040 & 0.053 &   & 0.069 & 0.054 & 0.048 &   & 0.076 & 0.051 & 0.053 \\ 
		2.2 &   & 0.051 & 0.048 & 0.054 &   & 0.059 & 0.040 & 0.061 &   & 0.057 & 0.048 & 0.058 \\ 
		3.1 &   & $\textbf{0.757}$ & $\textbf{0.821}$ & $\textbf{0.813}$ &   & $\textbf{0.824}$ & $\textbf{0.802}$ & $\textbf{0.815}$ &   & $\textbf{0.815}$ & $\textbf{0.801}$ & $\textbf{0.794}$ \\ 
		3.2 &   & $\textbf{0.779}$ & $\textbf{0.804}$ & $\textbf{0.838}$ &   & $\textbf{0.802}$ & $\textbf{0.811}$ & $\textbf{0.804}$ &   & $\textbf{0.836}$ & $\textbf{0.801}$ & $\textbf{0.815}$ \\ 
		4.1 &   & $\textbf{0.794}$ & $\textbf{0.795}$ & $\textbf{0.774}$ &   & $\textbf{0.805}$ & $\textbf{0.799}$ & $\textbf{0.810}$ &   & $\textbf{0.858}$ & $\textbf{0.781}$ & $\textbf{0.790}$ \\ 
		4.2 &   & $\textbf{0.759}$ & $\textbf{0.786}$ & $\textbf{0.771}$ &   & $\textbf{0.820}$ & $\textbf{0.792}$ & $\textbf{0.806}$ &   & $\textbf{0.815}$ & $\textbf{0.798}$ & $\textbf{0.792}$ \\ 
		\midrule 
		
		\multicolumn{13}{l}{uLR = Univariate logistic regression} \\
		\multicolumn{13}{l}{mB = Multivariate Bernoulli} \\
		\multicolumn{13}{l}{mLR = Multivariate logistic regression} \\
		\multicolumn{13}{l}{Bold-faced entries have effect sizes that should lead to a superiority conclusion} \\
		
		\bottomrule 
	\end{tabular}
\end{table}

\paragraph{Average treatment effects}
Probabilities to conclude superiority of average treatment effects are presented in Table \ref{tab:pReject.ATE.k2}.
Decisions resulted in appropriate Type I error rates around $.05$ for each of the posterior distribution types under a least favorable scenario of no effect (i.e., ES $1.1$, $1.2$, $2.1$, $2.2$ of Any and Compensatory rules) and the proportions of correct superiority conclusions (i.e., power) were close to the targeted $.80$ under a priori estimated sample sizes when the true effect was larger than zero (i.e., ES $3.1$, $3.2$, $4.1$, $4.2$ of Any and Compensatory rules). 
These results showcase that a priori computed sample sizes result in adequate statistical decisions.

In general, multivariate logistic regression (mLR) performed comparable to stratified multivariate analysis (mB) in the estimation of average treatment effects: Type I=error rates of mB were around $.05$ and statistical power was close to the targeted $.80$ as well. 
Compared to univariate logistic regression analysis (uLR), statistical power of multivariate logistic regression (mLR) appeared less sensitive to the correlation of the data. 
Effect sizes $3.1/3.2$ and $4.1/4.2$ under the Compensatory rule demonstrate most clearly how power of uLR increased when the correlation moved from negative to positive, with uncorrelated data reaching the targeted $.80$. 
For these conditions, the sample size which the uLR model was fitted on was smaller and larger respectively than needed for an analysis that assumes uncorrelated data.
The difference between uLR and mLR was relatively subtle however, which is in line with the pattern of required sample sizes in Table \ref{tab:SampleSizes.k2}.
This table shows that differences in required sample sizes for different correlations were relatively small under most data-generating mechanisms.
This implies that the effect of using an incorrect sample size on statistical power is relatively limited under the data-generating mechanisms in the simulation study, in contrast with the scenarios presented in Table \ref{tab:SampleSize_example}.

\paragraph{Conditional average treatment effects}
The results of conditional treatment effects in the subpopulations are presented in Table \ref{tab:pReject.CATE.k2}.
Similar to average treatment effects, Type I error rates were around the targeted $.05$ under the least favorable scenarios of no effect (ES $1.1, 1.2$ for Any and Compensatory rules) for all estimation methods. 
The proportion to conclude superiority correctly was above $.80$ in all scenarios with a sample size exceeding the required sample size for CATEs.
In the scenarios where the sample size for CATEs was lower than requirer ($4.1$ and $4.2$ for the Any and Compensatory rules and $2.2$ and $3.2$ for the All rule), the power was below $.80$.

A comparison of estimations methods for the continuous covariate revealed that multivariate logistic regression (mLR) was generally more powerful than the stratified multivariate analysis (mB) approach when the covariate was continuous.
These effects are prominent in t $2.2$ and $3.2$ (All rule) as well as ES $4.2$ (Any and Compensatory rules). 
The statistical power of stratified multivariate analysis (mB) and multivariate logistic regression analysis (mLR) did not differ for the discrete covariate, as demonstrated under ES $2.1$ and $3.1$ (All rule) as well as ES $4.1$ (Any and Compensatory rules).

\begin{table}[htbp]
	\small\sf\centering
	\caption{Proportions of superiority decisions for CATEs with two outcome variables by data-generating mechanism, correlation, and decision rule.} 
	\label{tab:pReject.CATE.k2}
	\begin{tabular}{lp{0.02cm}rrrp{0.02cm}rrrp{0.02cm}rrr}
		\toprule 
		& & \multicolumn{3}{c}{$\rho < 0$} & & \multicolumn{3}{c}{$\rho = 0$} & & \multicolumn{3}{c}{$\rho > 0$} \\
		\cmidrule(l){3-5} 
		\cmidrule(l){7-9} 
		\cmidrule(l){11-13} 
		ES  & & mB & mLR-S & mLR-V & & mB & mLR-S & mLR-V & & mB & mLR-S & mLR-V \\
		\cmidrule(l){3-13} 
		\multicolumn{13}{c}{Rule = All} \\
		\cmidrule(l){3-13} 
		1.1 &   & 0.002 & - & 0.000 &   & 0.006 & - & 0.001 &   & 0.009 & - & 0.004 \\ 
		1.2 &   & 0.000 & 0.000 & 0.001 &   & 0.004 & 0.002 & 0.004 &   & 0.007 & 0.003 & 0.004 \\ 
		2.1 &   & $\textbf{0.999}$ & - & $\textbf{0.997}$ &   & $\textbf{0.999}$ & - & $\textbf{0.998}$ &   & $\textbf{1.000}$ & - & $\textbf{0.999}$ \\ 
		2.2 &   & $\textbf{0.484}$ & $\textbf{0.873}$ & $\textbf{0.998}$ &   & $\textbf{0.537}$ & $\textbf{0.854}$ & $\textbf{1.000}$ &   & $\textbf{0.529}$ & $\textbf{0.880}$ & $\textbf{1.000}$ \\ 
		3.1 &   & $\textbf{1.000}$ & - & $\textbf{1.000}$ &   & $\textbf{1.000}$ & - & $\textbf{1.000}$ &   & $\textbf{1.000}$ & - & $\textbf{1.000}$ \\ 
		3.2 &   & $\textbf{0.790}$ & $\textbf{0.972}$ & $\textbf{1.000}$ &   & $\textbf{0.801}$ & $\textbf{0.979}$ & $\textbf{1.000}$ &   & $\textbf{0.804}$ & $\textbf{0.982}$ & $\textbf{1.000}$ \\ 
		4.1 &   & 0.050 & - & 0.040 &   & 0.042 & - & 0.036 &   & 0.045 & - & 0.048 \\ 
		4.2 &   & 0.051 & 0.045 & 0.054 &   & 0.052 & 0.053 & 0.059 &   & 0.046 & 0.056 & 0.060 \\ 
		\cmidrule(l){3-13} 
		\multicolumn{13}{c}{Rule = Any} \\
		\cmidrule(l){3-13} 
		1.1 &   & 0.054 & - & 0.050 &   & 0.064 & - & 0.039 &   & 0.051 & - & 0.052 \\ 
		1.2 &   & 0.053 & 0.038 & 0.054 &   & 0.057 & 0.055 & 0.056 &   & 0.063 & 0.048 & 0.048 \\ 
		2.1 &   & $\textbf{1.000}$ & - & $\textbf{1.000}$ &   & $\textbf{1.000}$ & - & $\textbf{1.000}$ &   & $\textbf{1.000}$ & - & $\textbf{1.000}$ \\ 
		2.2 &   & $\textbf{0.933}$ & $\textbf{1.000}$ & $\textbf{1.000}$ &   & $\textbf{0.913}$ & $\textbf{0.999}$ & $\textbf{1.000}$ &   & $\textbf{0.904}$ & $\textbf{0.999}$ & $\textbf{1.000}$ \\ 
		3.1 &   & $\textbf{1.000}$ & - & $\textbf{1.000}$ &   & $\textbf{1.000}$ & - & $\textbf{1.000}$ &   & $\textbf{1.000}$ & - & $\textbf{1.000}$ \\ 
		3.2 &   & $\textbf{0.932}$ & $\textbf{0.999}$ & $\textbf{1.000}$ &   & $\textbf{0.939}$ & $\textbf{0.998}$ & $\textbf{1.000}$ &   & $\textbf{0.899}$ & $\textbf{0.999}$ & $\textbf{1.000}$ \\ 
		4.1 &   & $\textbf{0.251}$ & - & $\textbf{0.266}$ &   & $\textbf{0.251}$ & - & $\textbf{0.242}$ &   & $\textbf{0.233}$ & - & $\textbf{0.230}$ \\ 
		4.2 &   & $\textbf{0.336}$ & $\textbf{0.508}$ & $\textbf{0.181}$ &   & $\textbf{0.305}$ & $\textbf{0.522}$ & $\textbf{0.183}$ &   & $\textbf{0.308}$ & $\textbf{0.512}$ & $\textbf{0.174}$ \\ 
		\cmidrule(l){3-13} 
		\multicolumn{13}{c}{Rule = Compensatory} \\
		\cmidrule(l){3-13} 
		1.1 &   & 0.061 & - & 0.047 &   & 0.076 & - & 0.033 &   & 0.048 & - & 0.039 \\ 
		1.2 &   & 0.040 & 0.040 & 0.043 &   & 0.062 & 0.057 & 0.056 &   & 0.057 & 0.046 & 0.048 \\ 
		2.1 &   & $\textbf{1.000}$ & - & $\textbf{1.000}$ &   & $\textbf{1.000}$ & - & $\textbf{1.000}$ &   & $\textbf{1.000}$ & - & $\textbf{1.000}$ \\ 
		2.2 &   & $\textbf{0.980}$ & $\textbf{1.000}$ & $\textbf{1.000}$ &   & $\textbf{0.969}$ & $\textbf{1.000}$ & $\textbf{1.000}$ &   & $\textbf{0.945}$ & $\textbf{0.999}$ & $\textbf{1.000}$ \\ 
		3.1 &   & $\textbf{1.000}$ & - & $\textbf{1.000}$ &   & $\textbf{1.000}$ & - & $\textbf{1.000}$ &   & $\textbf{1.000}$ & - & $\textbf{1.000}$ \\ 
		3.2 &   & $\textbf{0.951}$ & $\textbf{1.000}$ & $\textbf{1.000}$ &   & $\textbf{0.953}$ & $\textbf{1.000}$ & $\textbf{1.000}$ &   & $\textbf{0.945}$ & $\textbf{1.000}$ & $\textbf{1.000}$ \\ 
		4.1 &   & $\textbf{0.283}$ & - & $\textbf{0.326}$ &   & $\textbf{0.292}$ & - & $\textbf{0.319}$ &   & $\textbf{0.287}$ & - & $\textbf{0.316}$ \\ 
		4.2 &   & $\textbf{0.390}$ & $\textbf{0.504}$ & $\textbf{0.190}$ &   & $\textbf{0.354}$ & $\textbf{0.534}$ & $\textbf{0.183}$ &   & $\textbf{0.359}$ & $\textbf{0.537}$ & $\textbf{0.232}$ \\ 
		\midrule 
		
		\multicolumn{13}{l}{mB = Multivariate Bernoulli} \\
		\multicolumn{13}{l}{mLR-S = Multivariate logistic regression - sample} \\
		\multicolumn{13}{l}{mLR-V = Multivariate logistic regression - value} \\
		\multicolumn{13}{l}{Bold-faced entries have effect sizes that should lead to a superiority conclusion} \\
		
		\bottomrule
	\end{tabular}
\end{table}


\section{Illustration}\label{s:application}

We applied the proposed method to a subset of data from the $n = 19,435$ subjects from the International Stroke Trial \citep{ISTCG1997}. 
We selected participants who were alive after six months and were treated with either a combined treatment (Aspirin + medium / high-dose Heparin) or one of the single treatments (Aspirin only), resulting in a sample of $n = 5,657$ participants, of which $n_{H+A} = 1,859$ were in the Heparin + Aspirin group (treatment = $1$) and $n_{A} = 3,798$ subjects were in the Aspirin group (treatment = $0$). 
We fitted the model in Equation \ref{eq:psi_IST2} to compare the effects of the two treatments on a) recurrent stroke within 14 days (0 = no; 1 = yes) and b) dependency after six months (0 = no, 1 = yes) while taking systolic blood pressure of the subjects (\textit{Bp}) into account. 

\subsection{Method}

We applied the two procedures from Subsection \ref{s:heterogeneity} (fixed values and interval of the covariate) to assess the multivariate and weighted treatment differences in three different types of patient populations: 
\begin{enumerate}
	\item Average treatment effects in the trial population; 
	\item Conditional treatment effects in populations defined by a fixed value. 
	Patient populations were defined by six different values of blood pressure, specifically $1, 2, \text{ and } 3$ standard deviations below and above the mean. 
	\item Conditional treatment effects in populations defined by an interval. 
	Patient populations were defined by two different regions of blood pressure: $Bp < -1 \text{ SD}$ (Low), and $Bp > 1 \text{ SD}$ (High). 	
\end{enumerate}

\begin{figure}
	\centering
	\includegraphics[width=0.9\linewidth]{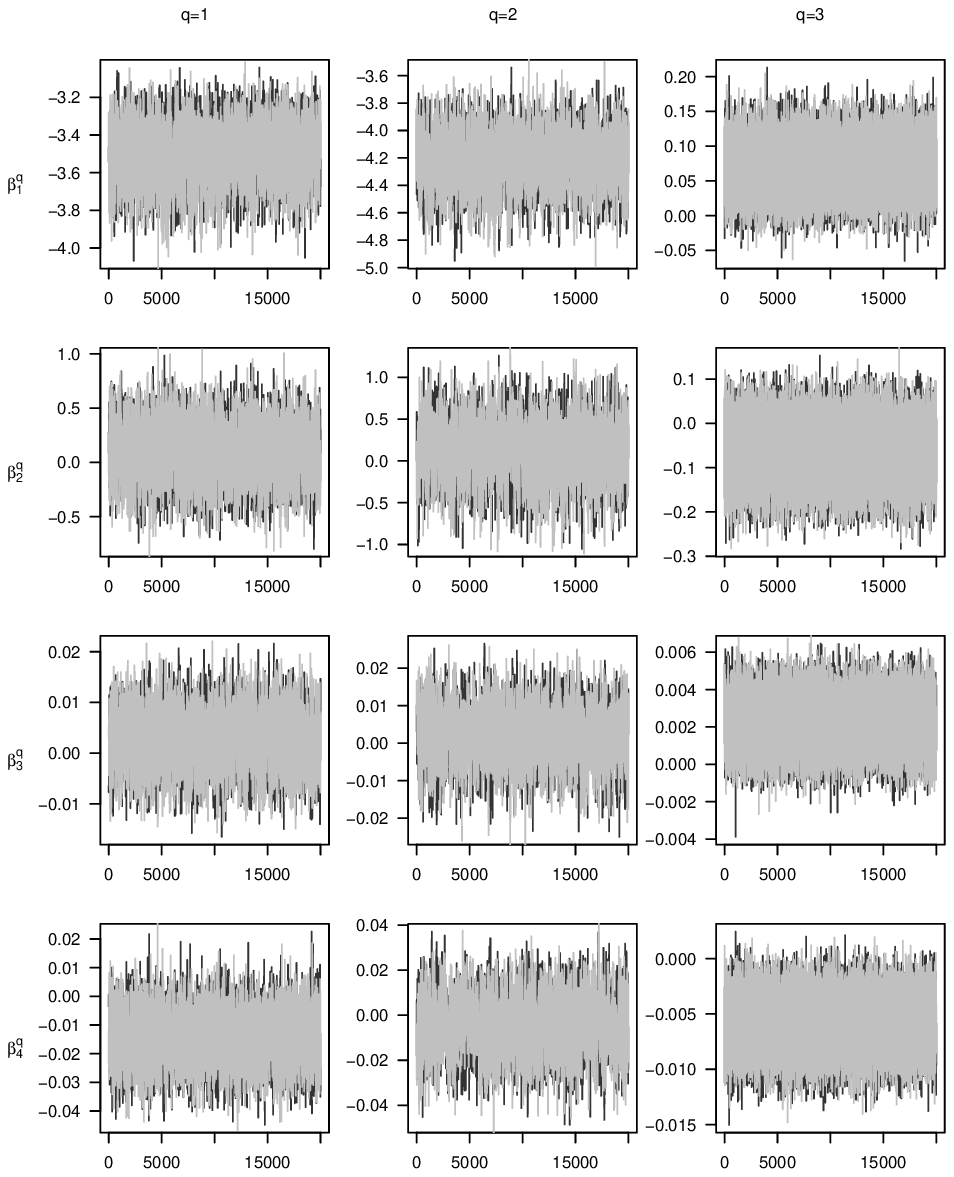}
	\caption{Traceplot of MCMC chains for the application of Bayesian multivariate logistic regression to the IST data}
	\label{fig:App_traceplot}
	
\end{figure}

Similar to the \nameref{s:evaluation}, we specified a diffuse multivariate normally distributed prior with means $\bm{b}^{q} = \bm{0}$ and variance matrix $\bm{B}^{0} = \text{diag}(1e^{1},\dots,1e^{1})$ for all regression coefficients, except the reference category ($strk = 0, dep = 0$). 
Prior covariances between regression coefficients were set at zero, implying that regression coefficients were independent a priori.
We ran three MCMC-chains via our proposed Gibbs sampler with $20,000$ iterations plus $10,000$ burnin iterations.
{
	Similar to the simulation study, we used traceplots and multivariate Gelman-Rubin convergence diagnostics to assess convergence \citep{Gelman1992,Brooks1998}. %
	Traceplots  (Figure \ref{fig:App_traceplot}) showed that chains mixed properly and the multivariate Gelman-Rubin convergence statistic had a value of $1.000$, implying that there were no signals of non-convergence. %
}\label{response:convergence_app}

We performed two-sided tests for the All, Any, and Compensatory rules. 
For the Compensatory rule, we assumed that long-term impaired functioning is more important than short-term complications and  specified weights $\bm{w} = (0.25,0.75)$ for recurring stroke in $14$ days and dependency at $6$ months respectively.
These weights implied that the longterm outcome was three times more relevant for the decision than the shortterm outcome. 
Since $\bm{\theta}_{T}$ reflects failure probabilities rather than success probabilities, the treatment is considered superior when there is sufficient evidence that the treatment difference of interest is \textit{smaller} than zero, while inferiority was concluded when the treatment difference of interest is \textit{larger} than zero. 
The two-sided test with a targeted Type I-error rate of $\alpha = .05$ was performed with a decision threshold $p_{cut} = 1 - \frac{\alpha}{2} = 0.975$ (Compensatory and All rules) and a for multiple tests corrected $p_{cut} = 1 - \frac{\alpha}{2K} = 0.9875 $ (Any rule).

\subsection{Results}

\begin{table}
	\small\sf\centering
	\caption{Average and conditional average treatment effects (ATE and CATE respectively) and their posterior probabilities (pp) in the IST data, by interval of blood pressure (Bp).
		Superiority or inferiority was concluded when $\bm{>}$ or $\bm{<}$ respectively.} 
	\label{tab:Application.Range_rbsp}
	\begin{tabular}{lrrrrrrrrr}
		\toprule
		Method  & & \multicolumn{1}{l}{$\bm{\delta}(\text{Bp})$} & \multicolumn{1}{l}{pp}  & Any & All & & \multicolumn{1}{l}{$\delta(\bm{w},\text{Bp})$} & \multicolumn{1}{l}{pp}  & Comp \\
		\midrule
		\multicolumn{4}{l}{\textbf{ATE} ($ - \infty < \text{Bp} < \infty$) 
		} & \multicolumn{6}{l}{$n_{\text{H+A}} = 1859$, $n_{\text{A}} = 3798$} \\
		\midrule
		mB &   & ( 0.005, -0.015) & (0.859, 0.151) & - & - &   & -0.010 & 0.182 & - \\ 
		mLR &   & ( 0.004, -0.014) & (0.825, 0.152) & - & - &   & -0.010 & 0.178 & - \\ 
		\midrule 
		\multicolumn{4}{l}{\textbf{CATE} ($ - \infty < \text{Bp} < -1 \text{ SD }$) 
		} & \multicolumn{6}{l}{$n_{\text{H+A}} = 316$, $n_{\text{A}} = 620$} \\
		\midrule 
		mB &   & (-0.001,  0.066) & (0.459, 0.972) & - & - &   &  0.049 & 0.970 & - \\ 
		mLR &   & ( 0.012,  0.043) & (0.932, 0.963) & - & - &   &  0.035 & 0.972 & - \\ 
		\midrule 
		\multicolumn{4}{l}{\textbf{CATE} ($+1 \text{ SD } < \text{Bp} < \infty$)
		} & \multicolumn{6}{l}{$n_{\text{H+A}} = 290$, $n_{\text{A}} = 646$} \\
		\midrule 
		mB &   & (-0.009, -0.052) & (0.214, 0.070) & - & - &   & -0.041 & 0.063 & - \\ 
		mLR &   & (-0.003, -0.081) & (0.330, 0.001) & $\bm{>}$ & - &   & -0.062 & 0.001 & $\bm{>}$ \\ 
		\midrule
		\multicolumn{10}{l}{mB = Multivariate Bernoulli analysis}\\
		\multicolumn{10}{l}{mLR= Multivariate logistic regression}\\
		\bottomrule
	\end{tabular}
\end{table}

\begin{table}
	\small\sf\centering
	\caption{Conditional average treatment effects in the IST data, by value of blood pressure (Bp).
		Superiority or inferiority was concluded when $\bm{>}$ or $\bm{<}$ respectively.
	} 
	\label{tab:Application.Value_rbsp}
	\begin{tabular}{lrrrrrrrrr}
		\toprule
		Value  & & \multicolumn{1}{l}{$\bm{\delta} (\text{Bp})$} & \multicolumn{1}{l}{pp}  & Any & All & & \multicolumn{1}{l}{$\delta(\bm{w}, \text{Bp})$} & \multicolumn{1}{l}{pp}  & Comp \\
		\midrule
		-3 SD &   & ( 0.029,  0.110) & (0.922, 0.994) & $\bm{<}$ & - &   &  0.090 & 0.996 & $\bm{<}$ \\ 
		-2 SD &   & ( 0.017,  0.068) & (0.930, 0.985) & - & - &   &  0.055 & 0.989 & $\bm{<}$ \\ 
		-1 SD &   & ( 0.009,  0.026) & (0.927, 0.908) & - & - &   &  0.022 & 0.929 & - \\ 
		+1 SD &   & (-0.001, -0.056) & (0.421, 0.002) & $\bm{>}$ & - &   & -0.042 & 0.002 & $\bm{>}$ \\ 
		+2 SD &   & (-0.004, -0.097) & (0.294, 0.001) & $\bm{>}$ & - &   & -0.074 & 0.001 & $\bm{>}$ \\ 
		+3 SD &   & (-0.007, -0.137) & (0.263, 0.001) & $\bm{>}$ & - &   & -0.104 & 0.001 & $\bm{>}$ \\ 
		\bottomrule
	\end{tabular}
\end{table}

Results are presented in Table \ref{tab:Application.Range_rbsp} for different intervals and in Table \ref{tab:Application.Value_rbsp} for fixed values of blood pressure.
Among the trial population, the regression-based and reference approaches resulted in similar treatment difference estimates and posterior probabilities.  
Treatment differences were close to zero and each of the decision rules resulted in the conclusion that it does not matter whether Aspirin was administered alone or in combination with Heparin.

These average treatment effects gave a limited impression of the efficacy of Aspirin and Heparin, since a picture of heterogeneous treatment effects emerged when conditional treatment effects among subpopulations were considered separately.
As opposed to Aspirin only, the combination of Aspirin and Heparin showed a trend towards higher failure probabilities on both dependent variables for patients with a lower blood pressure, while failure probabilities were generally lower among patients with a higher blood pressure.

\begin{figure}
	\centering
	\includegraphics[width=0.9\linewidth]{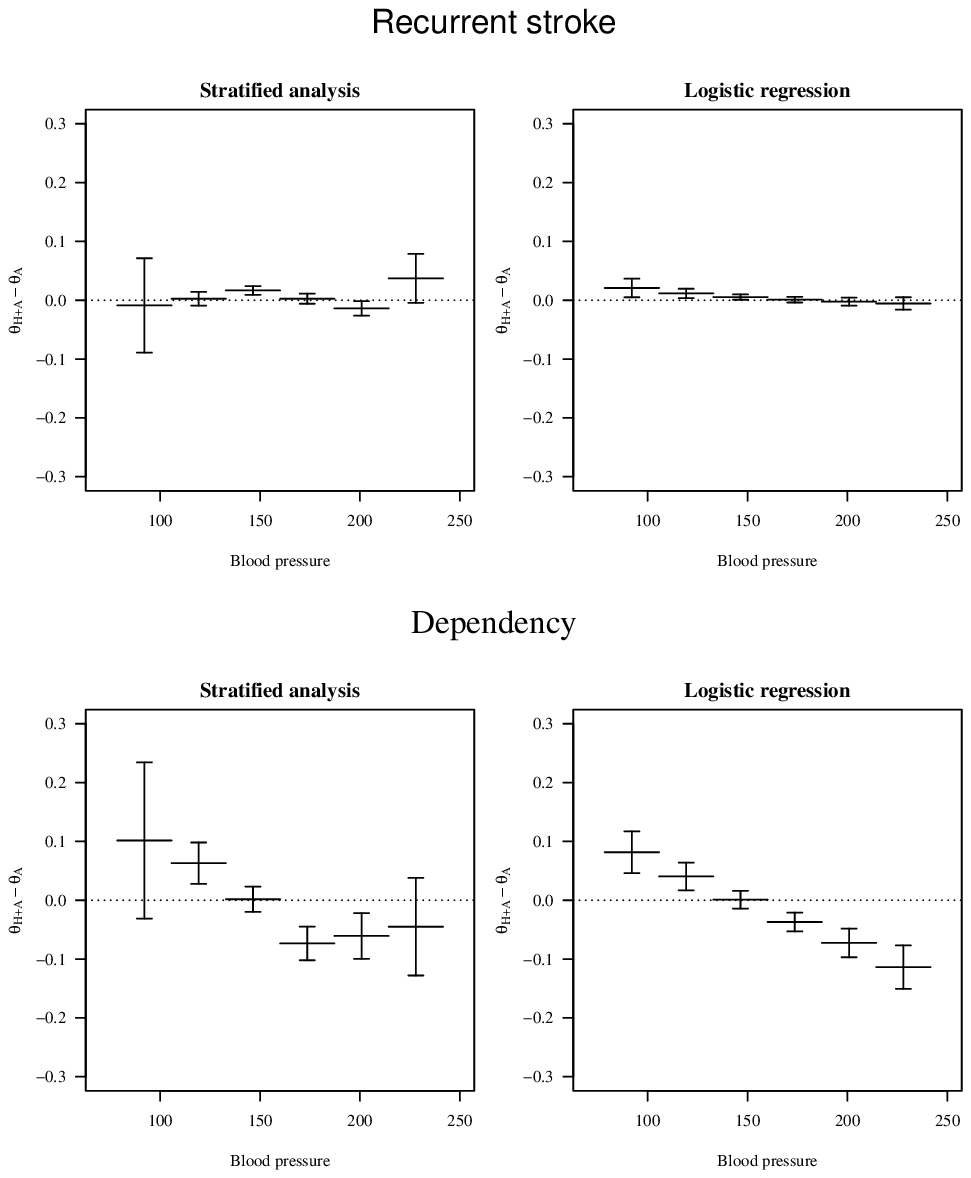}
	\caption{Comparison of CATEs and their standard deviations 
		per interval of blood pressure after stratified multivariate analysis (mB) and multivariate logistic regression (mLR). 
		Each interval reflects one standard deviation.}
	\label{fig:Application}
	
\end{figure}

A visual comparison of multivariate logistic regression (mLR) and stratified multivariate analysis (mB) of response data resulted in relatively similar estimates and posterior probabilities in the center of the distribution of blood pressure (e.g., between $-1$ SD and $+1$ SD), but deviated from the regression-based approach in the tails.
Point estimates of treatment differences demonstrated a less stable relation between blood pressure and treatment differences after stratification, as shown in Figure \ref{fig:Application}.
If the regression approach is flexible enough to properly model the effects over the full support of blood pressure, the different behavior in the tails of the covariate distribution might be explained by the smaller sample size after stratification, as implied by the larger error bars.

\section{Discussion}\label{s:discussion}

The current paper proposed a novel Bayesian multivariate logistic regression framework for analysis and decision-making with multiple correlated dependent variables. %
{
	The framework is suitable to capture treatment heterogeneity among (groups of) patients that are distinguishable by observed covariate information (i.e., conditional average treatment effects) and to estimate overall treatment effects among the full population (i.e., average treatment effects) under a wide range of scenarios. %
	In general, the proposed regression models were able to reproduce point estimates of average and conditional treatment differences correctly and resulted in decisions with anticipated error rates among the trial population and among subpopulations - as long as the sample was sufficiently large. %
	Further, anticipated decision error rates were found under a priori sample size estimation for different correlation structures (namely negatively correlated, uncorrelated, and positively correlated dependent variables) and for two- and three-dimensional dependent variables. %
	The illustration with the International Stroke Dataset demonstrated how conditional average treatment effects could provide a more in-depth understanding of results beyond average treatment effects. %

	Compared to other approaches, the Bayesian multivariate logistic regression framework showed favorable properties. %
	Decisions were more powerful than those obtained by multivariate stratified analysis when covariates were continuous, since they were based on information from the full sample rather than a subsample. %
	Moreover, the Bayesian multivariate logistic regression model was more effective in targeting statistical power compared to multiple univariate logistic regression analyses when the correlation between dependent variables was non-zero. %
	Whereas these effects were relatively subtle in the simulation study, the illustrative example in Section \ref{sss:samplesize_example} showcases that they are more prominent when correlations are further from zero. %
}\label{response:results}

{
	
	An advantage of the proposed multivariate logistic regression approach is its flexibility to model multivariate treatment effects with correlation structures that are free to vary over covariates, supporting accurate decision error rates and a priori sample size computations. 
	This flexiblity comes with additional parameters, compared to other multivariate logistic models for correlated binary dependent variables \citep[e.g.,][]{Malik1973,OBrien2004} and may result in computational issues when the number of parameters becomes too high.
	The Gibbs sampling procedure may become unstable when the sample size is too small compared to the number of parameters, although weakly informative priors may be helpful in stabilizing computations \citep{Gelman2008}. 
	Therefore, the model is most suitable for a limited number of dependent variables and (continous) covariates. 

	In practice, researchers are encouraged to consider model assumptions in real data. 
	Additional efforts may be undertaken to verify that the chosen generalized linear model fits the data well enough. 
	If the assumption of linearity on the log-odds scale does not hold, the modeling procedure may benefit from generalization to methods that are more flexible with respect to this assumption, such as (penalized) splines.
	Again, increased flexibility increases the number of parameters and should be balanced with a) the general risk of overfitting; and b) computational challenges as outlined above.
	In a more general sense, the researcher should determine which type of flexibility is most appropriate for the research question and data at hand.
	Further, researchers who aim to target decision error rates have to decide which treatment effect should be leading in the actual choice of sample size. 
	Under treatment heterogeneity, average and (multiple) conditional average treatment effects have different effect sizes by definition, resulting in different sample sizes and raising the question which considerations meaningfully guide this choice. 
	
	{
		Theoretically, the framework lends itself for use under a much wider range of scenarios than showcased in this paper. %
		Each of the elements - modeling, transformation, decision-making - can be replaced by an alternative, resulting in a large number of variations. %
		Some variations, such as a less computationally intensive analysis model, a wider range of prior distributions, and interim monitoring as an alternative to decision-making with a priori estimated sample sizes, were presented already \citep{Kavelaars2020}. %
		Here, we mention two additional suggestions to elaborate the framework. %
		First, in addition to the presented transformations to success probabilities and treatment,  transformations to other associations between treatment and outcome, such as relative risks and risk ratios, may be of interest and are worth investigating. %
		Second, other hypotheses than superiority and inferiority, such as non-inferiority or equality decision-making, can be relevant to be included in the framework as well \citep[see for a discussion][]{Ravenzwaaij2019}. %
		More flexible formulations of hypotheses and another perspective on the assessment of evidence can be achieved via the computation of Bayes factors \citep[see for an introduction e.g.,][]{Mulder2016}. %
	}\label{response:applicability}
	
	Other than the abovementioned variations, several directions for future research naturally follow from the current results. 
	First, the procedure theoretically lends itself for out-of-sample prediction to populations within or beyond the covariate range of the trial population. 
	The robustness of the framework in these applications remains to be investigated and may include evaluations of model fit.

	Second, research might shed light on further sample size considerations. 
	{The current paper provided tools to compute required sample sizes and to control decision error rates, if researchers are able to estimate effect sizes with reasonable accuracy prior to the study and when sample sizes are sufficiently large.}\label{response:magnitude}
	When sample sizes were relatively small, bias was introduced. 
	In line with our observations, small-sample bias in regression coefficients is a well-documented property of nonlinear regression methods in general \citep{Firth1993,Nemes2009}.  
	Although some bias in regression coefficients disappeared during transformation to joint response probabilities, success probabilities, and treatment differences, the mechanism is not yet fully understood.
	Hence, more light may be shed on circumstances for inheritance of distributional properties in the (non-linear) multinomial logistic transformation to obtain more elaborate insights into the minimum number of observations required for satisfactory model performance.
	Larger effect sizes (i.e., smaller sample sizes), complexity of the model (i.e., number of parameters), and events per variable are candidate factors to interact in their effects on model performance in small samples \citep{Jong2019}.
	There is no short answer to that question, but in practice power among different subpopulations might be balanced with the number of subjects a researcher is willing or able to include in the trial.
	Therefore, optimum sample sizes in these regression-based decision approaches remain to be investigated more elaborately.
	
	{
		Further, another interesting direction for future research would be to extend the proposed multivariate logistic regression model for estimating average and conditional average treatment effects and for decision-making with (discrete or continuous) latent variables to capture unexplained heterogeneity. %
		This extension falls outside of the scope of the current paper which focuses on modeling treatment heterogeneity caused by observed covariate information. %
	}\label{response:mixture2}

	Lastly, causal inference is less straightforward in (stratified) subgroup analysis as conditioning upon covariates might interfere with randomization \citep{EMA2019,FDA2019}. 
	Causal relationships might require additional checking of assumptions and tutorials by \citet{Hoogland2021} and \citet{Lipkovich2016} may be of help.

\newpage

\section*{Acknowledgements}
We thank anonymous reviewers for their valuable feedback on earlier drafts. 
Also, we thank The International Stroke Trial Collaborative Group for making the data from the second International Stroke Trial publicly available.

\section*{Funding}
The current work was supported by the Dutch Research Council (NWO) [no. 406.18.505].
The second International Stroke Trial was principally funded by the UK Medical Research Council, the UK Stroke Association, and the European Union BIOMED-1 program. Limited support for collaborators' meetings and travel was provided by Eli Lilly, Sterling Winthrop (now Bayer USA), Sanofi, and Bayer UK. 
Follow-up in Australia was supported by a grant from the National Heart Foundation and in Canada by a Nova Scotia Heart and Stroke Foundation grant. Czech Republic IST was supported by a grant from the IGA Ministry of Health. 
India IST was supported by the McMaster INCLEN program and the All India Institute of Medical Sciences. 
The IST in New Zealand was funded by the Julius Brendel Trust and the Lottery Grants Board. 
In Norway, the IST was supported by the Norwegian Council on Cardiovascular Disease and Nycomed (for insurance).

\section*{Declaration of interest}
The Authors declare that there is no conflict of interest.

\section*{Data availability}
The International Stroke Trial data that support the findings of this study are available 
with the identifier(s) [\url{http://doi.org/10.1186/1745-6215-12-101}].
The \texttt{R} code used to generate results in the \nameref{s:evaluation} and \nameref{s:application} sections can be found on GitHub \url{https://github.com/XynthiaKavelaars/Bayesian-multivariate-logistic-regression}.
\clearpage
\bibliographystyle{apalike}
\bibliography{Project2}

\begin{thebibliography}{}

\bibitem[Brooks and Gelman, 1998]{Brooks1998}
Brooks, S.~P. and Gelman, A. (1998).
\newblock General methods for monitoring convergence of iterative simulations.
\newblock {\em Journal of Computational and Graphical Statistics},
  7(4):434--455.

\bibitem[Chen and Ibrahim, 2000]{Chen2000}
Chen, M.-H. and Ibrahim, J.~G. (2000).
\newblock Power prior distributions for regression models.
\newblock {\em Statistical Science}, 15(1):46--60.

\bibitem[Chib, 1995]{Chib1995}
Chib, S. (1995).
\newblock Marginal likelihood from the gibbs output.
\newblock {\em Journal of the American Statistical Association},
  90(432):1313--1321.

\bibitem[Chow et~al., 2017]{Chow2017}
Chow, S.-C., Shao, J., Wang, H., and Lokhnygina, Y. (2017).
\newblock {\em Sample Size Calculations in Clinical Research: Third edition}.
\newblock Chapman and Hall/{CRC}.

\bibitem[Chuang-Stein et~al., 2006]{ChuangStein2006}
Chuang-Stein, C., Stryszak, P., Dmitrienko, A., and Offen, W. (2006).
\newblock Challenge of multiple co-primary endpoints: a new approach.
\newblock {\em Statistics in Medicine}, 26(6):1181--1192.

\bibitem[Dahl et~al., 2019]{Dahl2019}
Dahl, D.~B., Scott, D., Roosen, C., Magnusson, A., and Swinton, J. (2019).
\newblock {\em xtable: Export Tables to LaTeX or HTML}.
\newblock R package version 1.8-4.

\bibitem[Dai et~al., 2013]{Dai2013}
Dai, B., Ding, S., Wahba, G., et~al. (2013).
\newblock Multivariate bernoulli distribution.
\newblock {\em Bernoulli}, 19(4):1465--1483.

\bibitem[{De Jong} et~al., 2019]{Jong2019}
{De Jong}, V. M.~T., Eijkemans, M. J.~C., Calster, B., Timmerman, D., Moons, K.
  G.~M., Steyerberg, E.~W., and Smeden, M. (2019).
\newblock Sample size considerations and predictive performance of multinomial
  logistic prediction models.
\newblock {\em Statistics in Medicine}, 38(9):1601--1619.

\bibitem[{European Medicine Agency}, 2019]{EMA2019}
{European Medicine Agency} (2019).
\newblock {\em Guideline on the investigation of subgroups in confirmatory
  clinical trials}.

\bibitem[Firth, 1993]{Firth1993}
Firth, D. (1993).
\newblock Bias reduction of maximum likelihood estimates.
\newblock {\em Biometrika}, 80(1):27--38.

\bibitem[{Food and Drug Administration}, 2016]{FDA2016}
{Food and Drug Administration} (2016).
\newblock {\em Non-Inferiority Clinical Trials to Establish Effectiveness:
  Guidance for Industry}.

\bibitem[{Food and Drug Administration}, 2017]{FDA2017}
{Food and Drug Administration} (2017).
\newblock {\em Multiple Endpoints in Clinical Trials Guidance for Industry.}
\newblock Center for Biologics Evaluation and Research (CBER).

\bibitem[{Food and Drug Administration}, 2019]{FDA2019}
{Food and Drug Administration} (2019).
\newblock {\em Enrichment Strategies for Clinical Trials to Support
  Determination of Effectiveness of Human Drugs and Biological Products:
  Guidance for Industry.}
\newblock Center for Biologics Evaluation and Research (CBER).

\bibitem[Gelman et~al., 2008]{Gelman2008}
Gelman, A., Jakulin, A., Pittau, M.~G., and Su, Y.-S. (2008).
\newblock A weakly informative default prior distribution for logistic and
  other regression models.
\newblock {\em The Annals of Applied Statistics}, 2(4):1360--1383.

\bibitem[Gelman and Rubin, 1992]{Gelman1992}
Gelman, A. and Rubin, D.~B. (1992).
\newblock Inference from iterative simulation using multiple sequences.
\newblock {\em Statistical Science}, 7(4).

\bibitem[Goldberger and Buxton, 2013]{Goldberger2013}
Goldberger, J.~J. and Buxton, A.~E. (2013).
\newblock Personalized medicine vs guideline-based medicine.
\newblock {\em {JAMA}}, 309(24):2559.

\bibitem[Hamburg and Collins, 2010]{Hamburg2010}
Hamburg, M.~A. and Collins, F.~S. (2010).
\newblock The path to personalized medicine.
\newblock {\em New England Journal of Medicine}, 363(4):301--304.

\bibitem[Hoogland et~al., 2021]{Hoogland2021}
Hoogland, J., IntHout, J., Belias, M., Rovers, M.~M., Riley, R.~D., Jr, F.
  E.~H., Moons, K. G.~M., Debray, T. P.~A., and Reitsma, J.~B. (2021).
\newblock A tutorial on individualized treatment effect prediction from
  randomized trials with a binary endpoint.
\newblock {\em Statistics in Medicine}, 40(26):5961--5981.

\bibitem[{International Stroke Trial Collaborative Group}, 1997]{ISTCG1997}
{International Stroke Trial Collaborative Group} (1997).
\newblock The international stroke trial ({IST}): a randomised trial of
  aspirin, subcutaneous heparin, both, or neither among 19{\hphantom{,}}435
  patients with acute ischaemic stroke.
\newblock {\em The Lancet}, 349(9065):1569--1581.

\bibitem[Jones et~al., 2011]{Jones2011}
Jones, H.~E., Ohlssen, D.~I., Neuenschwander, B., Racine, A., and Branson, M.
  (2011).
\newblock Bayesian models for subgroup analysis in clinical trials.
\newblock {\em Clinical Trials}, 8(2):129--143.

\bibitem[Kavelaars et~al., 2020]{Kavelaars2020}
Kavelaars, X., Mulder, J., and Kaptein, M. (2020).
\newblock Decision-making with multiple correlated binary outcomes in clinical
  trials.
\newblock {\em Statistical Methods in Medical Research}, 29(11):3265--3277.

\bibitem[Kerman, 2011]{Kerman2011}
Kerman, J. (2011).
\newblock Neutral noninformative and informative conjugate beta and gamma prior
  distributions.
\newblock {\em Electronic Journal of Statistics}, 5(none).

\bibitem[Lipkovich et~al., 2016]{Lipkovich2016}
Lipkovich, I., Dmitrienko, A., and B., R. (2016).
\newblock Tutorial in biostatistics: data-driven subgroup identification and
  analysis in clinical trials.
\newblock {\em Statistics in Medicine}, 36(1):136--196.

\bibitem[Makalic and Schmidt, 2016]{Makalic2016}
Makalic, E. and Schmidt, D. (2016).
\newblock High-dimensional {B}ayesian regularised regression with the bayesreg
  package.
\newblock arXiv:1611.06649v3.

\bibitem[Malik and Abraham, 1973]{Malik1973}
Malik, H.~J. and Abraham, B. (1973).
\newblock Multivariate logistic distributions.
\newblock {\em The Annals of Statistics}, 1(3):588--590.

\bibitem[Marsman and Wagenmakers, 2016]{Marsman2016}
Marsman, M. and Wagenmakers, E.-J. (2016).
\newblock Three insights from a bayesian interpretation of the one-sided p
  value.
\newblock {\em Educational and Psychological Measurement}, 77(3):529--539.

\bibitem[Martin et~al., 2011]{Martin2011}
Martin, A.~D., Quinn, K.~M., and Park, J.~H. (2011).
\newblock {MCMCpack}: Markov chain monte carlo in {R}.
\newblock {\em Journal of Statistical Software}, 42(9):22.

\bibitem[McLachlan et~al., 2019]{McLachlan2019}
McLachlan, G.~J., Lee, S.~X., and Rathnayake, S.~I. (2019).
\newblock Finite mixture models.
\newblock {\em Annual Review of Statistics and Its Application}, 6(1):355--378.

\bibitem[Microsoft and Weston, 2020]{Microsoft2020a}
Microsoft and Weston, S. (2020).
\newblock {\em doParallel: Foreach Parallel Adaptor for the 'parallel'
  Package}.
\newblock R package version 1.0.16.

\bibitem[{Microsoft} and Weston, 2020]{Microsoft2020}
{Microsoft} and Weston, S. (2020).
\newblock {\em foreach: Provides Foreach Looping Construct}.
\newblock R package version 1.5.1.

\bibitem[Mulder and Wagenmakers, 2016]{Mulder2016}
Mulder, J. and Wagenmakers, E.-J. (2016).
\newblock Editors’ introduction to the special issue “bayes factors for
  testing hypotheses in psychological research: Practical relevance and new
  developments”.
\newblock {\em Journal of Mathematical Psychology}, 72:1--5.

\bibitem[Murray et~al., 2016]{Murray2016}
Murray, T.~A., Thall, P.~F., and Yuan, Y. (2016).
\newblock Utility-based designs for randomized comparative trials with
  categorical outcomes.
\newblock {\em Statistics in medicine}, 35(24):4285--4305.

\bibitem[Nemes et~al., 2009]{Nemes2009}
Nemes, S., Jonasson, J.~M., Genell, A., and Steineck, G. (2009).
\newblock Bias in odds ratios by logistic regression modelling and sample size.
\newblock {\em {BMC} Medical Research Methodology}, 9(1).

\bibitem[O'Brien, 1984]{OBrien1984}
O'Brien, P.~C. (1984).
\newblock Procedures for comparing samples with multiple endpoints.
\newblock {\em Biometrics}, pages 1079--1087.

\bibitem[O'Brien and Dunson, 2004]{OBrien2004}
O'Brien, S.~M. and Dunson, D.~B. (2004).
\newblock Bayesian multivariate logistic regression.
\newblock {\em Biometrics}, 60(3):739--746.

\bibitem[Olkin and Trikalinos, 2015]{Olkin2015}
Olkin, I. and Trikalinos, T.~A. (2015).
\newblock Constructions for a bivariate beta distribution.
\newblock {\em Statistics \& Probability Letters}, 96:54 -- 60.

\bibitem[Plummer et~al., 2006]{Plummer2006}
Plummer, M., Best, N., Cowles, K., and Vines, K. (2006).
\newblock Coda: Convergence diagnosis and output analysis for mcmc.
\newblock {\em R News}, 6(1):7--11.

\bibitem[Pocock, 1997]{Pocock1997}
Pocock, S.~J. (1997).
\newblock Clinical trials with multiple outcomes: a statistical perspective on
  their design, analysis, and interpretation.
\newblock {\em Controlled clinical trials}, 18(6):530--545.

\bibitem[Pocock et~al., 1987]{Pocock1987}
Pocock, S.~J., Geller, N.~L., and Tsiatis, A.~A. (1987).
\newblock The analysis of multiple endpoints in clinical trials.
\newblock {\em Biometrics}, 43(3):487.

\bibitem[Polson et~al., 2013]{polson2013}
Polson, N.~G., Scott, J.~G., and Windle, J. (2013).
\newblock Bayesian inference for logistic models using p{\'o}lya--gamma latent
  variables.
\newblock {\em Journal of the American statistical Association},
  108(504):1339--1349.

\bibitem[Prentice, 1997]{Prentice1997}
Prentice, R.~L. (1997).
\newblock Discussion: On the role and analysis of secondary outcomes in
  clinical trials.
\newblock {\em Controlled Clinical Trials}, 18(6):561--567.

\bibitem[{R Core Team}, 2020]{RCT2020}
{R Core Team} (2020).
\newblock {\em R: A Language and Environment for Statistical Computing}.
\newblock R Foundation for Statistical Computing, Vienna, Austria.

\bibitem[Ristl et~al., 2018]{Ristl2018}
Ristl, R., Urach, S., Rosenkranz, G., and Posch, M. (2018).
\newblock Methods for the analysis of multiple endpoints in small populations:
  A review.
\newblock {\em Journal of Biopharmaceutical Statistics}, 29(1):1--29.

\bibitem[Rossi et~al., 2005]{Rossi2005}
Rossi, P.~E., Allenby, G.~M., and McCulloch, R. (2005).
\newblock {\em Bayesian statistics and marketing}.
\newblock John Wiley \& Sons.

\bibitem[Sandercock et~al., 2011]{Sandercock2011}
Sandercock, P.~A., , Niewada, M., and Cz{\l}onkowska, A. (2011).
\newblock The international stroke trial database.
\newblock {\em Trials}, 12(1).

\bibitem[Senn and Bretz, 2007]{Senn2007}
Senn, S. and Bretz, F. (2007).
\newblock Power and sample size when multiple endpoints are considered.
\newblock {\em Pharmaceutical Statistics}, 6(3):161--170.

\bibitem[Simon, 2010]{Simon2010}
Simon, R. (2010).
\newblock Clinical trials for predictive medicine: new challenges and
  paradigms.
\newblock {\em Clinical Trials}, 7(5):516--524.

\bibitem[Sozu et~al., 2010]{Sozu2010}
Sozu, T., Sugimoto, T., and Hamasaki, T. (2010).
\newblock Sample size determination in clinical trials with multiple co-primary
  binary endpoints.
\newblock {\em Statistics in medicine}, 29:2169--2179.

\bibitem[Sozu et~al., 2016]{Sozu2016}
Sozu, T., Sugimoto, T., and Hamasaki, T. (2016).
\newblock Reducing unnecessary measurements in clinical trials with multiple
  primary endpoints.
\newblock {\em Journal of biopharmaceutical statistics}, 26(4):631--643.

\bibitem[Sullivan and Greenland, 2012]{Sullivan2012}
Sullivan, S.~G. and Greenland, S. (2012).
\newblock Bayesian regression in {SAS} software.
\newblock {\em International Journal of Epidemiology}, 42(1):308--317.

\bibitem[Tang et~al., 1993]{Tang1993}
Tang, D.-I., Geller, N.~L., and Pocock, S.~J. (1993).
\newblock On the design and analysis of randomized clinical trials with
  multiple endpoints.
\newblock {\em Biometrics}, 49(1):23.

\bibitem[Thall, 2020]{Thall2020}
Thall, P.~F. (2020).
\newblock Bayesian cancer clinical trial designs with subgroup-specific
  decisions.
\newblock {\em Contemporary Clinical Trials}, 90:105860.

\bibitem[Tuyl et~al., 2008]{Tuyl2008}
Tuyl, F., Gerlach, R., and Mengersen, K. (2008).
\newblock A comparison of bayes-laplace, jeffreys, and other priors: The case
  of zero events.
\newblock {\em The American Statistician}, 62(1):40--44.

\bibitem[{Van Ravenzwaaij} et~al., 2019]{Ravenzwaaij2019}
{Van Ravenzwaaij}, D., Monden, R., Tendeiro, J.~N., and Ioannidis, J. P.~A.
  (2019).
\newblock Bayes factors for superiority, non-inferiority, and equivalence
  designs.
\newblock {\em {BMC} Medical Research Methodology}, 19(1).

\bibitem[Venables and Ripley, 2002]{Venables2002}
Venables, W.~N. and Ripley, B.~D. (2002).
\newblock {\em Modern Applied Statistics with S}.
\newblock Springer, New York, fourth edition.
\newblock ISBN 0-387-95457-0.

\bibitem[Wang et~al., 2015]{Wang2015}
Wang, M., Spiegelman, D., Kuchiba, A., Lochhead, P., Kim, S., Chan, A.~T.,
  Poole, E.~M., Tamimi, R., Tworoger, S.~S., Giovannucci, E., Rosner, B., and
  Ogino, S. (2015).
\newblock Statistical methods for studying disease subtype heterogeneity.
\newblock {\em Statistics in Medicine}, 35(5):782--800.

\bibitem[Xiong et~al., 2005]{Xiong2005}
Xiong, C., Yu, K., Gao, F., Yan, Y., and Zhang, Z. (2005).
\newblock Power and sample size for clinical trials when efficacy is required
  in multiple endpoints: application to an alzheimer's treatment trial.
\newblock {\em Clinical Trials}, 2(5):387--393.

\bibitem[Yang et~al., 2021]{Yang2021}
Yang, S., Li, F., Thomas, L.~E., and Li, F. (2021).
\newblock Covariate adjustment in subgroup analyses of randomized clinical
  trials: A propensity score approach.
\newblock {\em Clinical Trials}, 18(5):570--581.

\bibitem[Zhao et~al., 2007]{Zhao2007}
Zhao, Y., Grambsch, P.~M., and Neaton, J.~D. (2007).
\newblock A decision rule for sequential monitoring of clinical trials with a
  primary and supportive outcome.
\newblock {\em Clinical Trials}, 4(2):140--153.

\end{thebibliography}

\newpage
\appendix
\section{Details of posterior computation}\label{app:posterior_computation}

The current section describes the Gibbs sampling procedure used to obtain parameters. 
To simplify notations, we omit the dependence on $\mathbf{x}$ in denoting functions that rely on covariates (e.g. $\mathbf{\phi}$, $\mathbf{\theta}$).

Starting from the likelihood of individual $K$-variate response $\mathbf{y}_{i}$ (Equation \ref{eq:psi2phi}), the likelihood of $n$ $K$-variate responses follows from taking the product over $n$ individual joint response probabilities in $Q$ response categories:
\begin{flalign}\label{eq:app_NH_mult_lik_beta}
	l(\mathbf{y}|\mathbf{\beta},\mathbf{x}) 
	= & 
	\prod_{i=1}^{n} 
	\prod_{q=1}^{Q-1}
	\left(\frac{\text{exp} \left[\psi^{q}_{i} 
		\right]}
	{\displaystyle\sum_{r=1}^{Q-1} \text{exp} \left[\psi^{r}_{i} 
		\right] + 1 }\right)^{I(\mathbf{y}_{i}=q)}
	\left(
	\frac{1}
	{\displaystyle\sum_{r=1}^{Q-1} \text{exp}\left[\psi^{r}_{i}
		\right] + 1}
	\right)^{I(\mathbf{y}_{i}=Q)}.
\end{flalign}

Following Polson et al. \citep{polson2013}, we introduce the P\'olya-gamma variable by rewriting the multivariate likelihood in Equation \ref{eq:app_NH_mult_lik_beta} as a series of binomial likelihoods.
The likelihood of $\mathbf{y}$ conditional on the parameters of the $q^{th}$ response category, $\mathbf{\beta}^{q}$, then equals:
\begin{flalign}\label{eq:app_NH_mult_lik_beta_bin}
	l(\mathbf{y}|\mathbf{\beta}^{q},\mathbf{\beta}^{-q})=
	\prod_{i=1}^{n} 
	\left(
	\frac{\text{exp}\left[\eta^{q}_{i}
		\right])}
	{\text{exp} \left[\eta^{q}_{i} 
		\right] + 1}
	\right)^{I(\mathbf{y}_{i}=q)}
	\left(
	\frac{1}{\text{exp} \left[\eta^{q}_{i} 
		\right] + 1}
	\right)^{1-I(\mathbf{y}_{i}=q)}  
\end{flalign}
\noindent where
$-q$ refers to all rows in $\mathbf{H}$ not having index $q$ and
$\eta^{q}_{i} 
=\psi^{q}_{i}
- \text{ln} \left(\displaystyle\displaystyle\sum_{m \neq\mathbf{H}_{q\cdot}} \text{exp} \left[\psi^{m}_{i} 
\right] \right).$
\\

The Polya-Gamma transformation to a Gaussian distribution relies on the following equality \citep{polson2013}:
\begin{flalign}\label{eq:app_NH_bin_lik_pg}
	\frac{\exp \left[\eta^{q}_{i}
		\right]}{\text{exp} \left[\eta^{q}_{i} 
		\right] + 1} & = 2 \text{exp} \left[(y_{i} - \frac{1}{2}) \eta^{q}_{i} 
	\right] \int_{0}^{\infty} \text{exp} \left[\frac{-\omega_{i} {\eta^{q}_{i} 
		}^{2}}{2}\right] p(\omega^{q}_{i}) d \omega^{q}_{i} 
\end{flalign}
\noindent where $\omega^{q}_{i}$ has a Polya-Gamma distribution, i.e. $p(\omega^{q}_{i}) \sim PG(1, \psi^{q}_{i}
)$.

If we use the equality in Equation \ref{eq:app_NH_bin_lik_pg}, the binomial likelihood in Equation \ref{eq:app_NH_mult_lik_beta_bin} can be transformed to a multivariate Gaussian likelihood by including an auxiliary P\'olya-Gamma variable $\omega^{q}_{i}$ \citep{polson2013}:

\begin{flalign}\label{eq:app_NH_mult_lik_pg}
	l(\mathbf{y}|\mathbf{\beta}^{q},\mathbf{\beta}^{-q}) & = \prod_{i=1}^{n} 
	\frac{\exp \left[\eta^{q}_{i} 
		\right]}{\text{exp} \left[\eta^{q}_{i} 
		\right]+ 1}\\\nonumber
	& = \prod_{i=1}^{n} 2 \text{exp} \left[(y_{i} - \frac{1}{2}) \eta^{q}_{i} 
	\right] \int_{0}^{\infty} \text{exp} \left[\frac{-\omega^{q}_{i} {\eta^{q}_{i} 
		}^{2}}{2}\right] p(\omega^{q}_{i}) d \omega^{q}_{i} \\\nonumber
	& = \prod_{i=1}^{n} \text{exp} \left[\kappa^{q}_{i} \omega^{q}_{i} \eta^{q}_{i} 
	- \frac{1}{2} (\eta^{q}_{i} 
	)^{2} \omega^{q}_{i}\right] PG(\omega^{q}_{i}|1,0) \\\nonumber
	& \propto \text{ exp} \left[ \frac{1}{2}(2 
	\mathbf{\kappa}^{q} \mathbf{\omega}^{q}
	\mathbf{\eta}^{q} 
	- \mathbf{\omega}^{q} (\mathbf{\eta}^{q} 
	)^{2} \right] \\\nonumber
	& \propto  \text{exp} \left[ - \frac{1}{2} (
	\mathbf{\kappa}^{q}
	- \mathbf{\eta}^{q})^{T} 
	\mathbf{\Omega}^{q}
	(\mathbf{\kappa}^{q}
	- \mathbf{\eta}^{q} 
	) \right]\\\nonumber
	& = \text{ exp} \left[- \frac{1}{2} (\mathbf{\kappa}^{q}
	- \mathbf{X} \mathbf{\beta}^{q} + \text{ln} [\displaystyle\sum_{m \neq q} \text{exp}(\mathbf{X} \mathbf{\beta}^{m})])^{T} 
	\mathbf{\Omega}^{q} 
	(\mathbf{\kappa}^{q}
	- \mathbf{X} \mathbf{\beta}^{q} + \text{ln} [\displaystyle\sum_{m \neq q} \text{exp}\left[\mathbf{X} \mathbf{\beta}^{m}\right]] \right]
	,\nonumber
\end{flalign}
\noindent where $\kappa^{q}_{i} = \frac{I(y_{i} = \mathbf{H}_{q\cdots}) - \frac{1}{2}}{\omega^{q}_{i}}$, $\mathbf{\kappa}^{q} = (\kappa^{q}_{1}, \dots, \kappa^{q}_{n})$, $\mathbf{\omega}^{q} = (\omega^{q}_{1},\dots,\omega^{q}_{n})$, and $\mathbf{\Omega}^{q} = \text{diag}(\mathbf{\omega}^{q})$.

\subsubsection{Prior distribution}

The Gaussian likelihood in Equation \ref{eq:app_NH_mult_lik_pg} is conditionally conjugate with a normal prior distribution on regression coefficients $\mathbf{\beta}^{q}$:
\begin{flalign}\label{eq:app_prior_beta}
	\mathbf{\beta}^{q} \sim N(\mathbf{b}^{q},\mathbf{\Beta}^{0q}) 
\end{flalign}  
\noindent where
$\mathbf{b}^{q}$ is the vector of prior means of regression coefficient vector $\mathbf{\beta}^{q}$ and
$\mathbf{\Beta}^{0q}$ is a $P \times P$ symmetric square matrix reflecting the prior precision of regression coefficients $\mathbf{\beta}^{q}$.
A researcher who is willing to include prior information regarding treatment effects into the analysis, has several options to specify prior hyperparameters for a normally distributed prior that is compatible with the Gibbs sampling procedure \citep[e.g.][]{Sullivan2012,Chen2000}. 
We discuss the specification of informative prior means $\mathbf{b}^{q}$ in terms of joint response probabilities $\mathbf{\phi}$ in the next Appendix. 

\subsubsection{Posterior distribution}

Bayesian statistical inference is done via the posterior distribution which is given by:
\begin{flalign}\label{eq:app_posterior}
	p(\mathbf{\beta}|\mathbf{y}) \propto & p(\mathbf{y}|\mathbf{\beta},\mathbf{x}) p(\mathbf{\beta}), 
\end{flalign}
The combination of a Polya-Gamma transformed Gaussian likelihood (Equation \ref{eq:app_NH_mult_lik_pg}) and a normal prior distribution (Equation \ref{eq:app_prior_beta}) respectively is proportional to a normally distributed posterior distribution, conditionally on Polya-Gamma variables in $\mathbf{\omega}^{q}$ \citep{polson2013}:

\begin{flalign}\label{eq:app_NH_mult_post}
	p(\mathbf{\beta}^{q}|\mathbf{Y}, \mathbf{\Omega}^{q}) \propto & p(\mathbf{y}|\mathbf{\beta}^{q}, \mathbf{\omega}^{q}) p(\mathbf{\beta}^{q})\\\nonumber
	\propto & \text{ exp} \left[- \frac{1}{2} (\mathbf{\kappa}^{q}
	- \mathbf{X} \mathbf{\beta}^{q} + \text{ln} [\displaystyle\sum_{m \neq q} \text{exp} \left[\mathbf{X} \mathbf{\beta}^{m}\right]])^{T} 
	\mathbf{\Omega}^{q} 
	(\mathbf{\kappa}^{q}
	- \mathbf{X} \mathbf{\beta}^{q} + \text{ln} [\displaystyle\sum_{m \neq q} \text{exp} \left[\mathbf{X} \mathbf{\beta}^{m}\right]]) \right] \times \\\nonumber
	& \text{ exp} \left[ - \frac{1}{2} (\mathbf{\beta}^{q} - \mathbf{b}^{q})^{T} (\mathbf{\Beta}^{q})^{-1} (\mathbf{\beta}^{q} - \mathbf{b}^{q}) \right] \\\nonumber
	\propto & N \left(\mathbf{V}^{q} ({\mathbf{X}}^{T} \mathbf{\Omega}^{q} (\mathbf{\kappa}^{q} + \text{ln} [\displaystyle\sum_{m \neq q} \text{exp} \left[\mathbf{X} \mathbf{\beta}^{m}\right]])
	+ (\mathbf{\Beta}^{q})^{-1}\mathbf{b}^{q}), 
	\mathbf{V}^{q} \right)
\end{flalign}
\noindent where 
$\mathbf{V}^{q} = ({\mathbf{X}}^{T} \mathbf{\Omega}^{q}  \mathbf{X} + (\mathbf{\Beta}^{q})^{-1})^{-1}$.
Similarly, subject-specific variable $\omega^{q}_{i}$ follows a Polya-Gamma distribution that depends on regression coefficients $\mathbf{\beta}^{q}$ via linear predictor $\psi^{q}_{i}$.

Updating these two conditional distributions via a Gibbs sampling procedure results in a sample from the posterior distribution of $\mathbf{\beta}$. 
Specifically, the sampling procedure involves iterating $L$ times over the following two steps for $q=1,\dots,Q-1$, while keeping $\mathbf{\beta}^{Q}$ fixed at zero:

\begin{enumerate}
	\item \label{enum_gibbs:sample_rcs}
	Draw a vector of $P + 1$ regression coefficients $\mathbf{\beta}^{q}|\mathbf{\omega}^{q}$ from a multivariate normal distribution with mean vector $\mathbf{m}^{q}$ and precision matrix $\mathbf{V}^{q}$.
	\begin{flalign}\label{eq:gibbs_1}
		&\mathbf{\beta}^{q}|\mathbf{\omega}^{q} \sim N(\mathbf{m}^{q},\mathbf{V}^{q})
	\end{flalign}
	\noindent%
	\begin{tabular}{lrl}
		where & $\lbrack \mathbf{V}^{q} \rbrack^{-1} = $ & $\mathbf{X} \mathbf{\Omega}^{q} \mathbf{X}+ \lbrack \mathbf{V}^{0q} \rbrack^{-1}$ \\
		& $\mathbf{m}^{q} = $ & $\mathbf{V}^{q}(\mathbf{X} (\mathbf{\kappa}^{q}+\mathbf{\Omega}^{q}
		\mathbf{c})+ \lbrack \mathbf{V}^{0q} \rbrack^{-1} \mathbf{m}^{0q})$ \\
		& $\mathbf{c} = $ & $ \left\{\text{ln} \left(\displaystyle\sum_{m \neq q} \text{exp} \left[\psi^{m}_{i} 
		\right] \right)_{i=1}^{n}\right\}. $ \\
	\end{tabular}
	
	\item \label{enum_gibbs:sample_omega}
	Sample $\mathbf{\omega}^{q}|\mathbf{\beta}^{q}$ as a vector of $n$ draws $\omega^{q}_{i}|\mathbf{\beta}^{q}$ from a P\'olya-Gamma distribution:
	\begin{flalign}\label{eq:gibbs_2}
		&\omega^{q}_{i}|\mathbf{\beta}^{q} \sim PG(1,\psi^{q}_{i} - \text{ln} \displaystyle\sum_{m \neq q} \text{exp} \left[\psi^{m}_{i} 
		\right]).
	\end{flalign}
\end{enumerate}
\noindent The Gibbs sampling procedure results in a sample of $L$ sets of regression coefficients from the posterior distribution of $\mathbf{\beta}$.

\newpage
\section{Specification of prior means of regression coefficients}\label{app:prior_specification}
In the current Section, we introduce a procedure to determine prior means, based on beliefs regarding success probabilities and correlations between them. 
We outline the procedure for two outcome variables and a linear predictor $\psi$ with one covariate and an interaction between the treatment and the covariate:
\begin{flalign}
	\psi^{q}_{T} & = \beta^{q}_{0} + \beta^{q}_{1} T + \beta^{q}_{2} x + \beta^{q}_{3} x \times T 
\end{flalign}

First, choose $x_{\text{L}}$ and $x_{\text{H}}$ as low and high values of covariate $x$ respectively. 
Next, specify success probabilities and correlations  $\mathbf{\theta}_{T} (x^{L})$, $\rho_{T} (x^{L})$, $\mathbf{\theta}_{T} (x^{H})$, and $\rho_{T} (x^{H})$ for each treatment $T$ that accompany the low and high values of covariates respectively.
These success probabilities $\mathbf{\theta}_{T} (x^{.})$ and correlations $\rho_{T} (x^{.})$ can be transformed to joint response probabilities $\mathbf{\phi}_{T} (x^{.})$ via the following set of equations:
\begin{flalign}
	\phi^{11}_{T}(x^{.}) & = \rho_{T} (x^{.}) \sqrt{\theta^{1}_{T} (x^{.}) 
		\left[1 - \theta^{1}_{T} (x^{.})\right] 
		\theta^{2}_{T}(x^{.}) 
		\left[1 - \theta^{2}_{T}(x^{.})\right]} + \theta^{1}_{T}(x^{.}) \theta^{2}_{T}(x^{.})\\\nonumber
	\phi^{10}_{T}(x^{.}) & = \theta^{1}_{T}(x^{.}) - \phi^{11}_{T}(x^{.})\\\nonumber
	\phi^{01}_{T}(x^{.}) & = \theta^{2}_{T}(x^{.}) - 
	\phi^{11}_{T}(x^{.})\\\nonumber
	\phi^{00}_{T}(x^{.}) & = 1 - \theta^{1}_{T}(x^{.}) - \theta^{2}_{T}(x^{.}) + \phi^{11}_{T}(x^{.})  \nonumber
\end{flalign}

For each response category $q$, joint responses $\phi^{q.}_{T}$ can be transformed to linear predictor $\psi^{q.}_{T}$ using the multinomial logistic link function in Equation \ref{eq:psi2phi}. 

Solving these linear predictors for $\mathbf{\beta}^{q}$ results in the following definitions of the elements in $\mathbf{\beta}^{q}$:
\begin{flalign}\label{eq:psi2beta}
	\beta^{q}_{0} & = 
	\frac{
		x^{\text{H}} \psi^{q}_{0} (x^{L}) - 
		x^{\text{L}} \psi^{q}_{0} (x^{H})}
	{x^{\text{H}} - x^{\text{L}}}  \\\nonumber
	\beta^{q}_{1} & = \frac{
		x^{\text{H}} \left[\psi^{q}_{1} (x^{L}) - \psi^{q}_{0} (x^{L})\right] + 
		x^{\text{L}} \left[\psi^{q}_{0} (x^{H}) - \psi^{q}_{1} (x^{H})\right]}
	{x^{\text{H}} - x^{\text{L}}}  \\\nonumber
	\beta^{q}_{2} & = \frac{\psi^{q}_{0} (x^{H} )- \psi^{q}_{0} (x^{L})}{x^{\text{H}} - x^{\text{L}}}  \\\nonumber
	\beta^{q}_{3} & = \frac{\psi^{q}_{1}(x^{H}) - \psi^{q}_{0} (x^{H}) - \psi^{q}_{1} (x^{L}) + \psi^{q}_{0} (x^{L})}{x^{\text{H}} - x^{\text{L}}}  
\end{flalign}

For example, if we would believe that treatment have the following parameters:
\begin{flalign*}
	\mathbf{\theta}^{L}_{1} &= (0.60, 0.70),  \rho^{L}_{1} = -0.30 &&\\\nonumber
	\mathbf{\theta}^{H}_{1} &= (0.40, 0.30),  \rho^{H}_{1} = -0.30&&\\\nonumber
	\mathbf{\theta}^{L}_{0} &= (0.40, 0.30),  \rho^{L}_{0} = -0.30&&\\\nonumber
	\mathbf{\theta}^{H}_{0} &= (0.60, 0.70),  \rho^{H}_{0} = -0.30,&&\nonumber
\end{flalign*} 
\noindent then the regression coefficients would be as presented in Table \ref{tab:appPriorMeans}.

\begin{table}
		\small\sf\centering
		\caption{Example of means of the prior distribution of regression coefficients}
		\label{tab:appPriorMeans}
		\begin{tabular}{rrrrr}
			\toprule
			& $q = 1$& $q = 2$& $q = 3$& $q = 4$\\ 
			\midrule
			$\beta^{q}_{0}$ & -0.000 & 0.766 & 0.766 & 0.000 \\ 
			$\beta^{q}_{1}$ & 0.000 & 0.000 & 0.000 & 0.000 \\ 
			$\beta^{q}_{2}$ & 1.902 & 0.781 & 1.121 & 0.000 \\ 
			$\beta^{q}_{3}$ & -3.804 & -1.562 & -2.241 & 0.000 \\ 
			\bottomrule
		\end{tabular}
\end{table}

\newpage
\section{Procedures for estimation and inference over a specified (sub)population}\label{app:alg_transformation}

\begin{algorithm}
	\caption{Transformation of posterior regression coefficients to posterior joint response probabilities based on fixed covariate values.}
	\label{alg:procedure_fixed_values}
	\begin{algorithmic}[1]
		\Statex Define $\mathbf{x} = x_{2},\dots,x_{P}$ as a vector of covariate values of interest
		\Statex Let $\mathbf{\beta}^{Q} = (0,\dots,0)$
		\For{draw $(l) \gets 1:L$}
		\For{treatment $T \gets 0:1$}
		\For{joint response $q \gets 1:Q$}
		\State Compute $\psi^{q(l)}_{T} 
		= \beta^{q(l)}_{0} + \beta^{q(l)}_{1} T + \beta^{q(l)}_{2} x + \beta^{q(l)}_{3} x \times T$		
		\State Compute $\phi^{q(l)}_{T} 
		= \frac{ \text{exp} \left[ \psi^{q(l)}_{T} 
			\right] }{\displaystyle\sum_{r=1}^{Q-1} \text{exp} \left[ \psi^{r(l)}_{T} 
			\right] + 1}$ 
		\EndFor
		\EndFor
		\EndFor
	\end{algorithmic}
\end{algorithm}

\begin{algorithm}
	\caption{Transformation of posterior regression coefficients to posterior joint response probabilities based on empirical marginalization.}\label{alg:procedure_empirical}
	\begin{algorithmic}[1]
		\Statex Let $\mathbf{\beta}^{Q} = (0,\dots,0)$
		
		\For{draw $(l) \gets 1:L$}
		\For{subject $i \gets 1:n$}
		\For{joint response $q \gets 1:Q$}
		\State Compute $\psi^{q(l)}_{i} 
		= \beta^{q(l)}_{1} T_{i} + \beta^{q(l)}_{2} x_{i} + \beta^{q(l)}_{3} x_{i} \times T_{i}$ 
		\State Compute $\phi^{q(l)}_{i} 
		= \frac{ \text{exp} \left[ \psi^{q(l)}_{i} 
			\right] }
		{\displaystyle\sum_{r=1}^{Q-1} \text{exp} \left[ \psi^{r(l)}_{i} 
			\right] + 1}$ 
		
		\For{$T \gets 0:1$}
		\State Compute $\mathbf{\phi}^{q(l)}_{T} (\mathbf{x}) = \frac{1}{\displaystyle\sum_{i=1}^{n} I(T_{i} = T)} \mathbf{\phi}^{q(l)}_{i} 
		I(T_{i} = T)$ 
		\EndFor
		\EndFor
		\EndFor
		\EndFor
	\end{algorithmic}
\end{algorithm}

\newpage
\section{Numerical evaluation with three outcome variables}\label{app:evaluation_k3}
In the current section, we present an evaluation of the BMLR framework with three dependent variables.

\subsection{Setup}

The evaluation largely follows the setup of the simulation with two dependent variables (Section \ref{s:evaluation}). 
Aspects that differ from this simulation will be discussed here. 

\subsubsection{Analysis}
We presented the results of Bayesian trivariate logistic regression analysis and compared it to a multivariate Bernoulli procedure.  

\subsubsection{Effect size}

We presented the results of the Bayesian trivariate logistic regression analysis for a selection of effect sizes, namely $1.1$ and $3.1$. 
Using the three correlation structures ($\rho < 0$, $\rho \approx 0$, and $\rho > 0$) for each of the effect sizes resulted in the six data generating mechanisms presented in Table \ref{tab:Delta.k3}.

\begin{table}[htbp]
	\small\sf\centering
	\caption{Parameters of average treatment effects (ATEs) in the trial and conditional average treatment effects (CATEs) in a subpopulation for tree outcome variables. } 
	\label{tab:Delta.k3}
	\begin{tabular}{llp{0.02cm}rrrp{0.02cm}rrr}
		\toprule
		
		&  & & \multicolumn{3}{c}{ATE} & & \multicolumn{3}{c}{CATE} \\ 
		\cmidrule(lr){3-6} 
		\cmidrule(l){7-10} 
		ES &  &  & \multicolumn{1}{l}{$(\delta_{1}, \delta_{2}, \delta_{3})$} & \multicolumn{1}{l}{$\delta (\bm{\mathsf{w}})$} & \multicolumn{1}{l}{$\rho_{\theta^{k},\theta^{l}}$} &  & \multicolumn{1}{l}{$(\delta_{1}, \delta_{2}, \delta_{3})$} & \multicolumn{1}{l}{$\delta (\bm{\mathsf{w}})$} & \multicolumn{1}{l}{$\rho_{\theta^{k},\theta^{l}}$} \\
		\midrule
		1.1 & D &   & (0.000,  0.000,  0.000) &  0.000 & -0.160 &   & (0.000,  0.000,  0.000) &  0.000 & -0.200 \\ 
		&  &   &   &   &  0.030 &   &   &   &  0.000 \\ 
		&  &   &   &   &  0.220 &   &   &   &  0.200 \\ 
		3.1 & D &   & (0.100,  0.000,  0.100) &  0.075 & -0.152 &   & (0.300,  0.200,  0.300) &  0.275 & -0.200 \\ 
		&  &   &   &   &  0.040 &   &   &   &  0.000 \\ 
		&  &   &   &   &  0.232 &   &   &   &  0.200 \\ 
		\midrule 
		\multicolumn{10}{l}{ES = Effect size, D = Discrete covariate, C = Continuous covariate} \\
		\bottomrule
	\end{tabular}
\end{table}

\subsubsection{Sample size}
Similar to the \nameref{s:evaluation}, we applied the All, Any, and Compensatory rules. We assigned the Compensatory rule unequal weights $\bm{w} = (0.50, 0.25, 0.25)$.

The required sample sizes for three outcome variables are computed via the procedure described in Section \ref{ss:sample_size}, targeting at a Type I error rate of $.05$ and a power of $.80$. 
The sample sizes are presented in Table \ref{tab:SampleSizes.k3}.

\begin{table}[htbp]
	\small\sf\centering
	\caption{Required sample sizes to evaluate the average treatment effect (ATE) and conditional treatment effect (CATE) for three outcome variables.} 
	\label{tab:SampleSizes.k3}
	\begin{tabular}{llp{0.02cm}rrrp{0.02cm}rrrp{0.02cm}rrr}
		\toprule
		&  & & \multicolumn{3}{c}{All} & & \multicolumn{3}{c}{Any} & & \multicolumn{3}{c}{Compensatory} \\
		
		\cmidrule(lr){4-6}
		\cmidrule(l){8-10}
		\cmidrule(l){12-14} ES & $\rho_{k^{\theta},l^{\theta}}$  & & ATE & CATE & Sub & & ATE & CATE & Sub & & ATE & CATE & Sub \\
		\midrule
		1.1 & $<0$ &   & - & - & 500 &   & - & - & 500 &   & - & - & 500 \\ 
		& $\approx 0$ &   & - & - & 500 &   & - & - & 500 &   & - & - & 500 \\ 
		& $>0$ &   & - & - & 500 &   & - & - & 500 &   & - & - & 500 \\ 
		3.1 & $<0$ &   & - & 79 & 500 &   & 234 & 20 & 117 &   & 153 & 9 & 77 \\ 
		& $\approx 0$ &   & - & 79 & 500 &   & 255 & 23 & 128 &   & 218 & 14 & 109 \\ 
		& $>0$ &   & - & 78 & 500 &   & 276 & 26 & 138 &   & 284 & 19 & 142 \\ 
		\midrule 
		
		\multicolumn{14}{l}{Sub = expected size of subsample} \\ 
		\bottomrule
	\end{tabular}
\end{table}

\subsubsection{Decision rule}
We performed a right-sided (superiority) test aiming at a Type I-error rate of $\alpha = .05$.
We used a decision threshold $p_{cut} = 1 - \alpha = 0.95$ (Compensatory and All rules) and a for multiple tests corrected $p_{cut} = 1 - \frac{\alpha}{K} = 0.981$ (Any rule) \citep{Marsman2016, Kavelaars2020, Sozu2016}.

\subsubsection{Procedure} 
To stabilize computations, we used $20,000$ iterations for the multivariate Bernoulli model. 

\subsection{Results}
\begin{table}[htbp]
	\small\sf\centering
	\caption{Proportions of superiority decisions for three outcome variables by data-generating mechanism, correlation, and decision rule.} 
	\label{tab:pReject.k3.RS}
	\begin{tabular}{llp{0.02cm}rrp{0.02cm}rrp{0.02cm}rr}
		\toprule 
		& & & \multicolumn{2}{c}{$\rho < 0$} & & \multicolumn{2}{c}{$\rho = 0$} & & \multicolumn{2}{c}{$\rho > 0$} \\
		\cmidrule(l){4-5} 
		\cmidrule(l){7-8} 
		\cmidrule(l){10-11} 
		ES & Type & & mB & mLR & & mB & mLR & & mB & mLR \\
		\cmidrule(l){4-11} 
		\multicolumn{11}{c}{Rule = All} \\
		\cmidrule(l){4-11} 
		1.1 &  ATE &   & 0.000 & 0.000 &   & 0.000 & 0.001 &   & 0.004 & 0.002 \\ 
		3.1 &  ATE &   & 0.045 & 0.045 &   & 0.058 & 0.048 &   & 0.044 & 0.058 \\ 
		1.1 &  CATE &   & 0.000 & 0.000 &   & 0.000 & 0.000 &   & 0.001 & 0.000 \\ 
		3.1 &  CATE &   & $\textbf{1.000}$ & $\textbf{1.000}$ &   & $\textbf{1.000}$ & $\textbf{1.000}$ &   & $\textbf{1.000}$ & $\textbf{1.000}$ \\ 
		\cmidrule(l){4-11} 
		\multicolumn{11}{c}{Rule = Any} \\
		\cmidrule(l){4-11} 
		1.1 &  ATE &   & 0.049 & 0.056 &   & 0.045 & 0.050 &   & 0.046 & 0.056 \\ 
		3.1 &  ATE &   & $\textbf{0.814}$ & $\textbf{0.822}$ &   & $\textbf{0.796}$ & $\textbf{0.775}$ &   & $\textbf{0.815}$ & $\textbf{0.775}$ \\ 
		1.1 &  CATE &   & 0.047 & 0.050 &   & 0.042 & 0.037 &   & 0.063 & 0.032 \\ 
		3.1 &  CATE &   & $\textbf{1.000}$ & $\textbf{1.000}$ &   & $\textbf{1.000}$ & $\textbf{1.000}$ &   & $\textbf{1.000}$ & $\textbf{1.000}$ \\ 
		\cmidrule(l){4-11} 
		\multicolumn{11}{c}{Rule = Compensatory} \\
		\cmidrule(l){4-11} 
		1.1 &  ATE &   & 0.048 & 0.068 &   & 0.050 & 0.052 &   & 0.052 & 0.063 \\ 
		3.1 &  ATE &   & $\textbf{0.781}$ & $\textbf{0.826}$ &   & $\textbf{0.788}$ & $\textbf{0.757}$ &   & $\textbf{0.787}$ & $\textbf{0.776}$ \\ 
		1.1 &  CATE &   & 0.051 & 0.043 &   & 0.056 & 0.029 &   & 0.053 & 0.035 \\ 
		3.1 &  CATE &   & $\textbf{1.000}$ & $\textbf{1.000}$ &   & $\textbf{1.000}$ & $\textbf{1.000}$ &   & $\textbf{1.000}$ & $\textbf{1.000}$ \\ 
		\midrule 
		
		\multicolumn{11}{l}{mB = Multivariate Bernoulli} \\
		\multicolumn{11}{l}{mLR = Multivariate logistic regression} \\
		\multicolumn{11}{l}{Bold-faced entries should lead to a superiority conclusion} \\
		
		\bottomrule 
	\end{tabular}
\end{table}

\end{document}